\documentclass[preprint]{elsarticle}
\usepackage{lineno,hyperref}
\usepackage{amsmath}
\usepackage{xspace}
\usepackage{siunitx}
\usepackage{graphicx}
\usepackage{booktabs}
\usepackage{makecell}
\usepackage{subcaption}
\usepackage{color}
\usepackage{placeins}
\modulolinenumbers[5]
\journal{Nuclear Instruments and Methods A}
\newcommand{\microns}{\si{\micro\meter}\xspace}
\begin{document}
\begin{frontmatter}
\title{Spatial resolution studies using point spread function extraction in optically read out Micromegas and GEM detectors}

%in prepration
\author{A. Cools}
\author{E. Ferrer-Ribas}
\author{T. Papaevangelou}
\author{E. C. Pollacco}
\address{IRFU, CEA, Universit\'e Paris-Saclay, F-91191 Gif-sur-Yvette, France}

%in prepration
\author{M.~Lisowska}
\author{F.M.~Brunbauer}
\author{E.~Oliveri}
\address{European Organization for Nuclear Research (CERN), CH-1211 Geneve 23, Switzerland}

\author{F.J.~Iguaz}
\address{SOLEIL Synchrotron, L’Orme des Merisiers, Départementale 128, 91190 Saint-Aubin, France}

%\author[a]{F.M.~Brunbauer}
%\author[a,c]{K.J.~Flöthner} 
%\author[g]{F.~Garcia} 
%author[a]{D.~Janssens} 
%\author[a,d]{M.~Lisowska}
%\author[a,c]{H.~Müller} %Hans.Muller@cern.ch
%\author[a]{E.~Oliveri} %Eraldo.Oliveri@cern.ch
%\author[a,e]{G.~Orlandini} %Eraldo.Oliveri@cern.ch
%\author[a,b]{D.~Pfeiffer} %Dorothea.Pfeiffer@cern.ch
%\author[a]{L.~Ropelewski} %Leszek.Ropelewski@cern.ch
%\author[a]{F.~Sauli}
%\author[b]{J.~Samarati} %Jerome.Samarati@cern.ch
%\author[a]{L.~Scharenberg} %lucian.scharenberg@cern.ch
%author[a]{M.~van Stenis} %Miranda.van.Stenis@cern.ch
%\author[a]{R.~Veenhof}

%\cortext[cor]{Corresponding author}

%\affiliation[a]{European Organization for Nuclear Research (CERN), CH-1211 Geneve 23, Switzerland}
%\affiliation[b]{European Spallation Source (ESS ERIC), P.O. Box 176, SE-22100 Lund, Sweden}
%\affiliation[c]{University of Bonn, Regina-Pacis-Weg 3, 53113 Bonn, Germany}
%\affiliation[d]{Institut de Recherche sur les lois Fondamentales de l’Univers (IRFU, CEA), Université Paris Saclay, F-91191 Gif-sur-Yvette, France}
%\affiliation[e]{Friedrich-Alexander-Universität Erlangen-Nürnberg, Schloßplatz 4, 91054 Erlangen, Germany}
%\affiliation[g]{Helsinki Institute of Physics, University of Helsinki, Finland}


\begin{abstract}

Optically read out gaseous detectors are used in track reconstruction and imaging applications requiring high granularity images. %pixellated readout.
Among resolution-determining factors, the amplification stage plays a crucial role and optimisations of detector geometry are pursued to maximise spatial resolution. To compare MicroPattern Gaseous Detector (MPGD) technologies, focused low-energy X-ray beams at the SOLEIL synchrotron facility were used to record and extract point spread function widths with Micromegas and GEM detectors. Point spread function width of $\approx$108\,\microns for Micromegas and  $\approx$127\,\microns for GEM foils were extracted. The scanning of the beam with different intensities, energies and across the detector active region can be used to quantify resolution-limiting factors and improve imaging detectors using MPGD amplification stages.

\end{abstract}


\end{frontmatter}
%\linenumbers
%\maketitle
\newpage
\tableofcontents
\newpage
%%%%%%%%%%%%%%%%%%%%%%%%%%%%%%%%%%%%%%%%%%%%%%%%%%%%%%
\section{Introduction}
\label{section:Introduction}
%%%%%%%%%%%%%%%%%%%%%%%%%%%%%%%%%%%%%%%%%%%%%%%%%%%%%%
Sample imaging is crucial for understanding the underlying phenomena in fields such as biology~\cite{BERG2022106195}, medicine~\cite{10.1371/journal.pone.0009470}, materials science~\cite{LEHMANN2011161, Lehmann_2011}, and even art~\cite{ALFELD201781}. Additionally, visual inspection in industrial production is crucial for safety, security screening and longevity of machinery and electronics. As a result, probes that utilise X-ray, $\beta$, neutron (\textit{n}), or proton (\textit{p}) beams with integrated cameras have become standard tools.

In this paper we report  on camera developments based  on Micro Pattern Gaseous Detectors (MPGDs) technologies. The detectors use gas as the conversion volume where electrons produced in the initial ionisation are transported onto planar gas amplifiers (MPGDs). 

MPGDs can provide excellent spatial and good charge resolution~\cite{Attie:2021fbp, SCHARENBERG2021165576, Kudryavtsev_2020}, by inducing charge on a segmented  anode. With appropriate electronics and  digitization~\cite{Baron:2011,Baron:2017kld}, images are assembled in real time. Key advantages of this structure include the following: the system can be used with several  beams (\textit{n}, \textit{p},  $\beta$ and X-Rays) with only minor modifications for \textit{n} and hence versatile; the exposed samples can be made relatively large (tens of cm per dimension); on-line images show acceptable to good image contrast depending on irradiation intensity and attenuation factors. Importantly, the setup is radiation hard and therefore compatible with industrial use and radiation therapy. However,  a major drawback is the significant cost and complexity of the electronics  when the channel count exceeds several 10$^3$.

Low-emittance X-ray beams at the SOLEIL synchrotron facility at various energies with beam dimensions down to $20\times20\,\si{\micro\meter}^2$ were used to provide highly parallel incident beams. The measurement configurations are described in Section~\ref{section:Measurements}. The collected data allowed to extract Point Spread Function (PSF) widths under different detector geometries and MPGD parameter settings.  The PSF analysis (Section~\ref{sec:results}) explores optimal PSF widths reaching down to 110\,\microns. The PSF analysis with different deconvoluted components derives the intrinsic spatial resolution yielding a tentative value of in the range of 70\,\microns. Furthermore, recorded images enable to observe electron transport processes via photons emitted in the Micromegas avalanche for the first time, for instance electron deflection around the pillars. Finally, in Section~\ref{section:Conclusion}  we underline that the results reached to date can be improved through structural changes of the camera and photon transport and recording.  

%%%%%%%%%%%%%%%%%%%%%%%%%%%%%%%%%%%%%%%%%%%%%%%%%%%%%%
\section{Experimental setup description}
\label{section:Setup}
%%%%%%%%%%%%%%%%%%%%%%%%%%%%%%%%%%%%%%%%%%%%%%%%%%%%%%%
\subsection{Metrology beam line at SOLEIL and setup description}
The detectors have been installed in the Metrology beamline\cite{Metrology1, Metrology2} at the SOLEIL synchroton facility that produces a beam of hard X-rays with small divergence and high flux covering energies from 6\,keV to 28\,keV. The beam was shaped by focusing mirrors and collimating slits. Beam sizes from 20\,$\times$ 20\,$\microns^2$ to 1\,$\times$1\,mm$^2$ have been explored. The beam energy can be tuned by a monochromator. 
While tests were predominantly performed at 6\,keV, exploration measurements at higher energies such as 18\,keV and 28\,keV were conducted. Beam fluxes of 9.4$\times10^9$\,ph/s/mm$^2$ were measured with a photodiode at the detector plane position. 

\begin{figure}[htb!]
    \centering
    \includegraphics[width=0.58\textwidth]{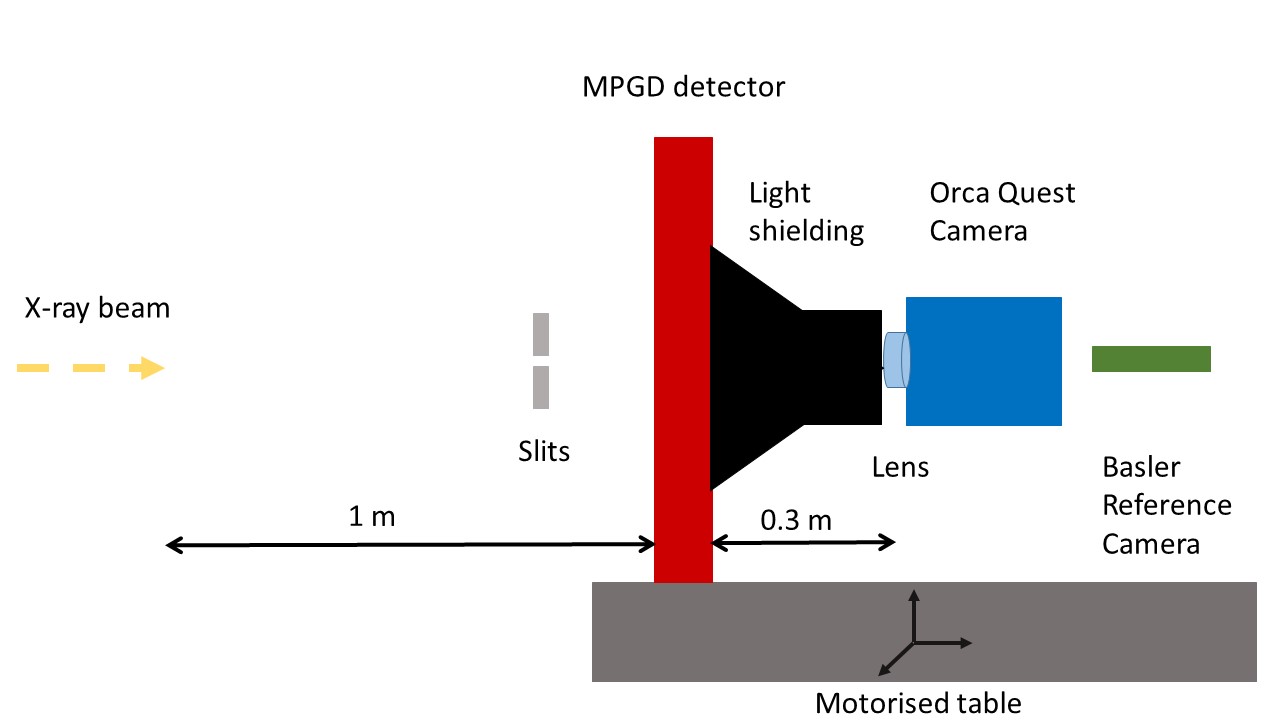}
    \includegraphics[width=0.40\textwidth]{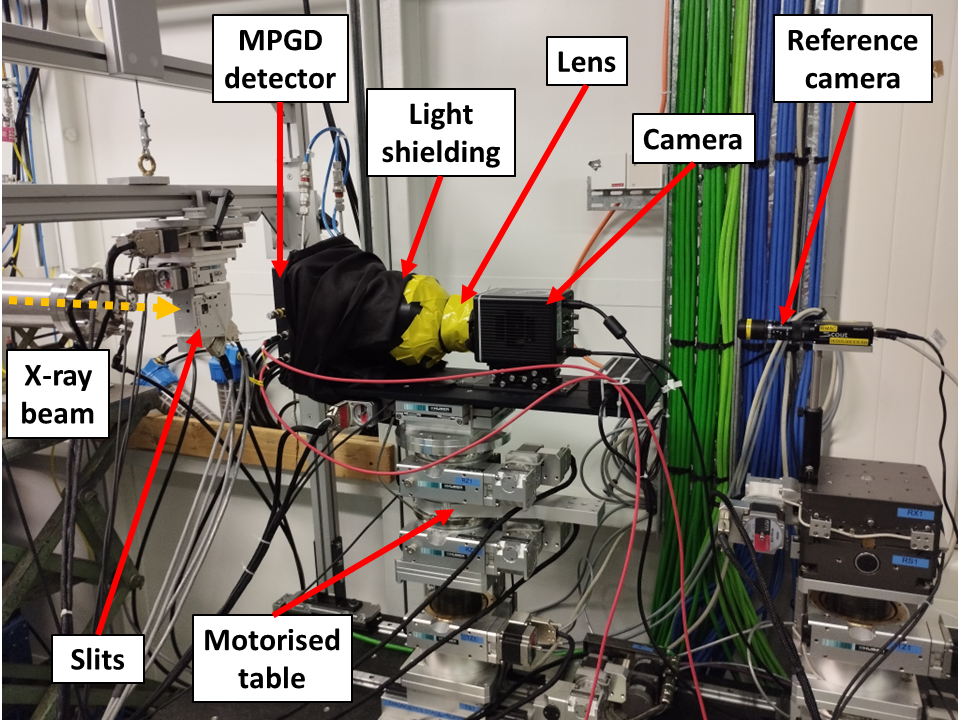}
\caption{Left: Sketch of the experimental set-up giving an overview of the X-ray beam with the optically read out MPGD detector under test on a motorised table allowing three-dimensional movement. Right: View of the experimental setup.}
    \label{SOLEIL-set-up}
\end{figure}

\begin{figure}[htb!]
    \centering
    \includegraphics[width=0.45\textwidth]{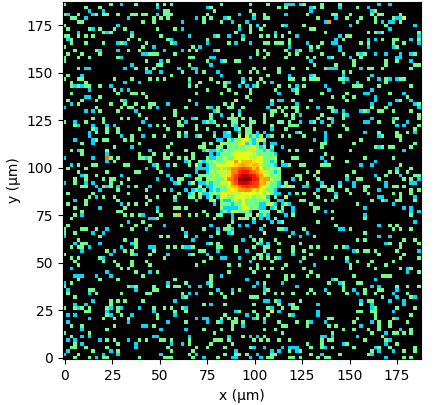}      
    \caption{Basler camera pixel intensity frame of a 6\,keV 20 X-rays beam. The beam size is 20 $\times$20$\microns^2$.}
\label{Fig:BaslerCam}
\end{figure}

The detector system, comprising an MPGD detector, a lens, and a camera, is mounted on a motorized table capable of both horizontal and vertical adjustments. A sketch and a view of the experimental set-up is shown in Figure~\ref{SOLEIL-set-up}. A Basler camera~\cite{AltaVision} positioned behind the detector serves to record 2D beam-profile images. By pairing the camera with a lens with a magnification factor of 2 and a scintillating crystal, a granularity of 1.88 $\times$ 1.88~$\microns^2$ was reached as shown in Figure~\ref{Fig:BaslerCam}. 

A Hamamatsu ORCA®-Quest CMOS camera~\cite{Hamamatsu} is placed behind the MPGD detector. The camera features a quantum efficiency of about 80\,$\%$ at the gas mixture main scintillation emission wavelength with a 9.4\,megapixel granularity and a readout noise as low as 0.27 electrons RMS. Two different lenses with focal lengths of 25\,mm and 50\,mm were employed. 

\subsection{Micromegas}
\begin{figure}
   \centering
   \includegraphics[width=0.9\textwidth]{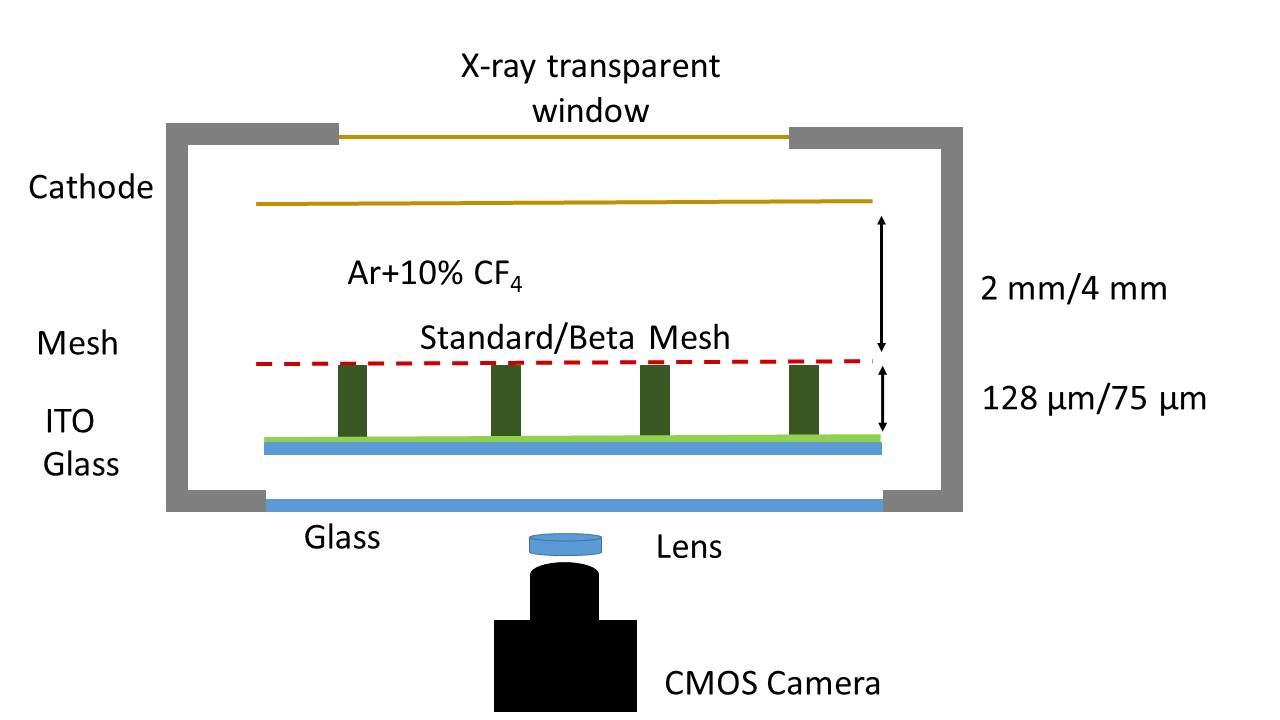}
   \caption{Schematic of the Micromegas optical readout detector. An aluminium window and a kapton cathode transparent to X-rays were employed. A CMOS camera was used for imaging.}
    \label{Micromegas-description}
\end{figure}

Glass Micromegas detectors~\cite{BRUNBAUER2020163320} with an active area of 8$\times$8\,cm$^2$ were mounted in a test chamber with a thin transparent Kapton entrance X-ray window. Drift gaps of 2\,mm and
4\,mm were used. The amplification gap, defined by insulating 500\,$\microns$-diameter pillars,  measured 128\,\microns for a standard 
woven stainless-steel mesh (45\,\microns  aperture, 18\,\microns wire widths and 30\,\microns thickness)
and 75\,\microns for an electroformed thinner mesh (18\,\microns thickness), dubbed Beta mesh. Microscope images of the two types of meshes are shown in Figure~\ref{Fig:MPGDGeometries} (a, b). To polarise the detector, the mesh is grounded and positive high voltage is applied to the Indium Tin Oxide (ITO) coating on a Quartz plate used as anode. The detector is operated at gains of several 10$^2$-10$^3$ in an Argon-10\%CF$_{4}$ gas mixture. 
The camera is placed behind the Micromegas glass anode (5\,mm thick quartz coated with an ITO layer) and a second 5\,mm thick quartz glass piece without ITO, which forms the gas-tight vessel as shown in Figure~\ref{Micromegas-description}.

\subsection{GEM}
\label{GEMSetup}
\begin{figure}[htb!]
    \centering
    \includegraphics[width=1\textwidth]{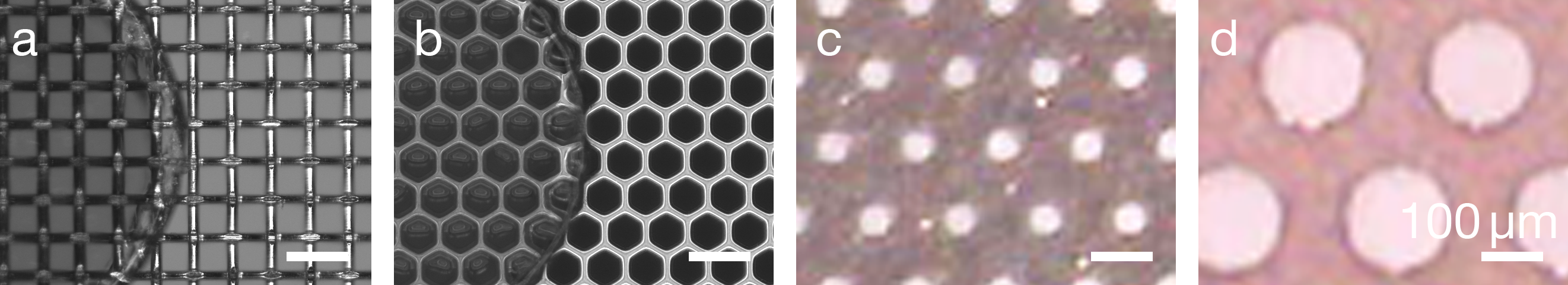}    
    \caption{Glass Micromegas with standard woven mesh (a) and electroformed mesh (b). Hole pattern of standard thin GEM (c) and glass GEM (d). The scale is the same for all images.}
    \label{Fig:MPGDGeometries}
\end{figure}

GEM detectors with 10\,cm x 10\,cm active area with two different hole geometries were evaluated as shown in Figure \ref{Fig:MPGDGeometries} (c, d). Standard thin GEM foils feature 70 \microns diameter holes with a pitch of 140\,\microns structured in a 50 \microns thick polyimide foil with 5\,\microns thick copper electrodes~\cite{Sauli:1997qp}. Glass GEM foils are produced by etching of photoetchable glass~\cite{glassGEM} and feature 160-180 \microns diameter holes with a pitch of 280\,\microns structured in a 570 \microns thick glass substrate with 2\,\microns thick copper electrodes. 

Detectors were assembled in a dedicated gas vessel placed behind collimator slits as shown in Figure \ref{SOLEIL-set-up} and read out with the same CMOS camera as described above. Single GEMs were used and a copper drift electrode was placed to define a 2\,mm thick drift region. No anode was placed below the GEM and the amplification current was collected at the bottom electrode of the GEM.

The bottom electrode of the GEM was grounded and negative high voltage was applied to the top GEM electrode and the cathode to define drift and amplification fields. GEMs were operated at gain values of several hundreds with a maximum gain of $10^3$. For glass GEMs, a high voltage of 1270\,V to 1370\,V was applied, for standard thin GEMs a high voltage of 450\,V to 510\,V was used. A drift field strength of 350\,V/cm was applied.

%%%%%%%%%%%%%%%%%%%%%%%%%%%%%%%%%%%%%%%%%%%%%%%%%%%%%%
\section{Measurements description}
\label{section:Measurements}
%%%%%%%%%%%%%%%%%%%%%%%%%%%%%%%%%%%%%%%%%%%%%%%%%%%%%%
Various configurations of MPGDs were investigated with X-ray beams with the same collimator and entrance window fixed. For the Micromegas detector, two amplification gap sizes (75\,$\microns$ and 128\,$\microns$) and two drift lengths (2 and 4\,mm) were probed. For the GEM detectors, the standard thin GEM foils and glass GEM foils described in section~\ref{section:Setup} were tested. Optical elements were also explored, with two different magnification lenses (0.1 and 1) and aperture sizes ranging from f/0.95 to f/2.8. Additionally, we assessed the impact of a mirror on the optical axis. The beam size was adjusted from 20 $\times$20$\,\microns^2$  to 1$\times$1\,mm$^2$, allowing illumination of different detector regions by altering the detector's position. Energies ranging from 6\,keV to 28\,keV were tested. A summary of the different configurations tested is given in Table~\ref{tab:MPGD-tested-parameters}.

\begin{table}[]
    \centering
    \begin{tabular}{|c|c|}
    \hline
        Micromegas & \\
         Mesh type &  Standard / Beta mesh\\
         Amplification gap &128\,\microns / 75\,\microns \\
         Drift gap & 2\,mm / 4\,mm \\
         Beam Energy & 6\,keV / 18\,keV / 28 \,keV\\
        % Beam Size & 20\,$\times$ 20\,$\microns^2$ to 1\,$\times$1\,mm$^2$\\
          \hline
          \hline
          GEM type & Standard thin GEM / Glass GEM\\
         Hole diameter & 70 $\microns$ / 160 $\microns$\\
         Hole pitch & 140 $\microns$ / 280 $\microns$\\
         Drift gap & 2\,mm \\
         %Lens & Magnification 0.1 and 1\\
          %  & f/0.95 to f/2.8 \\
        
         Beam Energy & 6\,keV\\
         \hline
         \hline
          Camera & ORCA Quest\\
          \hline
          \hline
          Lens & Magnification 0.1 / 1\\
            & f/0.95 to f/2.8 \\
            \hline
            \hline
            Beam Size & 20\,$\times$ 20\,$\microns^2$ to 1\,$\times$1\,mm$^2$\\
            \hline
    \end{tabular}
    \caption{Summary table of the different configurations tested with the Micromegas and the GEM set-ups.}
    \label{tab:MPGD-tested-parameters}
\end{table}

%\begin{table}[]
%    \centering
%    \begin{tabular}{|c|c|}
%    \hline
%         GEM type & Standard thin GEM / Glass GEM\\
%         Hole diameter & 70 $\microns$ / 160 $\microns$\\
%         Hole pitch & 140 $\microns$ / 280 $\microns$\\
%         Drift gap & 2\,mm \\
%         Lens & Magnification 0.1 and 1\\
%            & f/0.95 to f/2.8 \\
%         Camera & ORCA Quest\\
%         Beam Energy & 6\,keV\\
%         Beam Size & 20\,$\times$ 20\,$\microns^2$ to 1\,$\times$1\,mm$^2$\\
%          \hline
%    \end{tabular}
%    \caption{Summary table of the different configurations tested with the GEM setup.}
%    \label{tab:GEM-tested-parameters}
%\end{table}

%%%%%%%%%%%%
\subsection*{Image Acquisition and Processing}
%%%%%%%%%%%%

Beam images were acquired with 30\,s exposure time and 10 frames per setting were recorded and averaged. Background images with the beam in off mode were recorded for each configuration with the same parameters. Averaged background images were subtracted from averaged signal images removing possible ambient light contributions. The intensity of primary scintillation light was negligible and could not be observed with these recording parameters. Image processing was performed using ImageJ software~\cite{Schneider2012}.

%%%%%%%%%%%%%%%%%%%%%%%%%%%%%%%%%%%%%%%%%%%%%%%%%%%%%%
\section{Results}
%%%%%%%%%%%%%%%%%%%%%%%%%%%%%%%%%%%%%%%%%%%%%%%%%%%%%%
\label{sec:results}
The detector's response was assessed by Point Spread Function (PSF) method, a common technique in imaging analysis. The PSF characterizes how an imaging system responds to a point source, such as a collimated point-like X-ray beam in the present study. While an ideal sensor would yield a PSF that is infinitely small and sharply peaked, real sensors produce blurred images due to various physical and optical effects. In the investigated MPGD detectors, the mean free path of photoelectrons, transverse electron diffusion and avalanche multiplication in the amplification structures contribute to blurring, alongside optical phenomena such as lens effects and reflections within the system. In this section, we detail the measured PSFs for the detector configurations described in Section~\ref{section:Measurements}.
%%%%%%%%%%%%
\subsection{PSF measurements with Micromegas}
%%%%%%%%%%%%
%%
\subsubsection*{Beam description as a function of beam detector position} 
\begin{figure}[htb!]
    \centering
    \includegraphics[width=\textwidth]{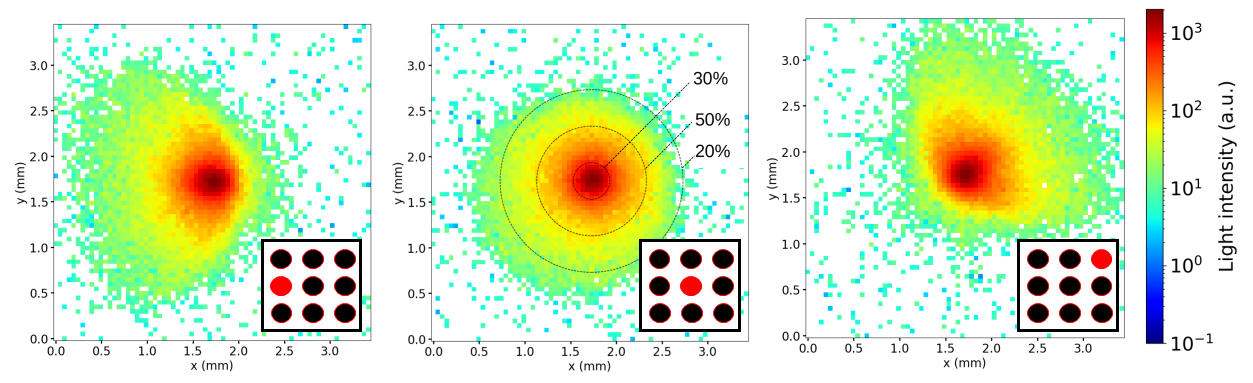}      
    \caption{2D profiles of a point-like beam located at different positions in the detector as shown  in the insert. The lens aperture is f/0.95, the drift field is 350\,V/cm and the amplification field is 47\,kV/cm. The optical aberrations with an asymmetric shape are visible. The light intensity scale is logarithmic, to visualise easily the total light response across several orders of magnitude.}
\label{Fig:PSFPositions}
\end{figure}

To assess the detector's uniformity, 9 different detector positions spaced 2\,cm apart were illuminated with a 20$\times20\,\microns^2$ beam. A schematic of these positions is depicted in the inserts of Figure~\ref{Fig:PSFPositions}. The beta mesh was evaluated with a lens aperture set to f/0.95, using 5\,s exposure time frames for 1\,min acquisition time. As depicted in Figure~\ref{Fig:PSFPositions}, the detector response  exhibits varying profiles depending the beam's location within the detector. Notably, the signal exhibits a blur directed away from the optical axis of the system, a phenomenon known as optical aberration coma, which increases quadratically with the distance of the source from the optical axis.

%\begin{figure}[htpb]
%    \centering
%    \includegraphics[width=0.4\textwidth]{./figures/BeamPositions.png}      
%    \caption{Sketch of the beam positions on the detector. The red spots represent the beam and their position labels are represented. The distance between the positions is 2\,cm.}
%\label{Fig:BeamPositions}
%\end{figure}

%\begin{figure}[htpb]
%    \centering
%    \includegraphics[width=0.29\textwidth]{./figures/Soleil2606_70_350_21_6_30_0.95.png} 
%    \includegraphics[width=0.29\textwidth]{./figures/Soleil2606_70_350_22_6_30_0.95.png} 
%    \includegraphics[width=0.29\textwidth]{./figures/Soleil2606_70_350_13_6_30_0.95.png}      
%    \includegraphics[width=0.08\textwidth]{./figures/Soleil2606_70_350_22_6_30_0.95_colorbar.png}      
 %   \caption{2D profiles of a point-like beam located at the center left-21 (left), center-22 (center) and top right-13 (right) of the image plane. Percentage of the total light amount among several distances from the beam axis (center). The lens aperture is f/0.95, the drift field is 350\,V/cm and the amplification field is 47\,kV/cm. The optical aberrations with an asymmetric shape are visible. The light intensity scale is logarithmic, facilitating visualization of the total light response across several orders of magnitude, spanning from the PSF center to its periphery. Here we show the ratio of light detected at various distances from the PSF center, revealing that approximately 80\,\% of the total light is captured within a 600\,\microns radius around the beam center.}
%\label{Fig:PSFPositions}
%\end{figure}

\begin{figure}[htb!]
    \centering
    \includegraphics[width=\textwidth]{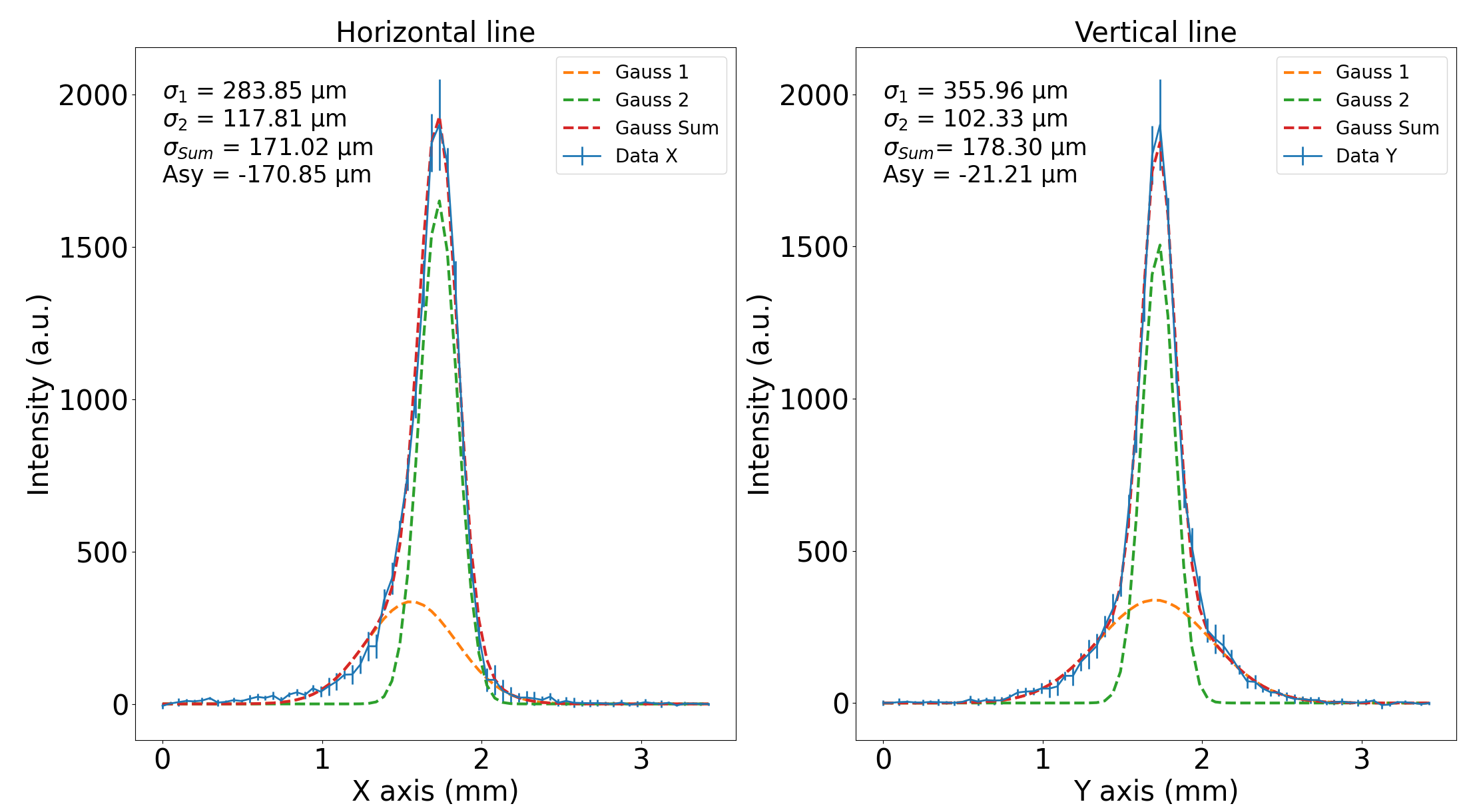}  
    \caption{1D horizontal and vertical line profiles of the PSF displayed in Figure\,\ref{Fig:PSFPositions} at the center left beam position. On the horizontal profile (left), the Gaussian distribution 1 (orange line) is shifted to the second one (green line). This shift is implemented in the variable $Asy$ which is the difference between the first and the second distributions mean values. The lens aperture is f/0.95, the drift field is 350\,V/cm and the amplification field is 47\,kV/cm.}
\label{Fig:1DLinesBeam21}
\end{figure}

A method has been developed to extract the spatial resolution from the PSF and facilitate the comparison of PSF profiles across configurations. This involves generating the vertical and the horizontal 1D profiles from the images and conducting appropriate fitting. Figure~\ref{Fig:1DLinesBeam21} shows that the horizontal 1D profile typically exhibits a Gaussian shape with an asymmetric component indicative of spatial resolution degradation due to aberrations. To account for this asymmetry, a fitting function detailed in Equation~\ref{eq:2a} is employed, incorporating two Gaussian distributions with parameters $\mu_1$, $\mu_2$, $\sigma_1$ and $\sigma_2$ (mean values and the standard deviations, respectively) and the ratio of amplitudes of the two Gaussian functions, $c$. The spatial resolution is commonly defined by the standard deviation of the distribution, expressed by Equation~\ref{eq:2b} for a distribution comprising a sum of $N$ Gaussian functions. In our case, a simplified distribution with $N = 2$ is adopted, as outlined in Equation~\ref{eq:2c}.

\begin{equation}\label{eq:2a} 
\begin{centering}
f=A \cdot \left(c \cdot \exp\left(-\frac{{(x - \mu_1)^2}}{{2 \cdot \sigma_1^2}}\right) + (1-c) \cdot \exp\left(-\frac{{(x - \mu_2)^2}}{{2 \cdot \sigma_2^2}}\right)\right)
\end{centering}
\end{equation}

\begin{equation}\label{eq:2b} 
\begin{centering}
\sigma^2=\sum_{i=1}^{N} c_i^2 \cdot \sigma_i^2 + \sum_{i=1}^{N-1}\sum_{j=i+1}^{N} c_i \cdot c_j \cdot (\sigma_i^2+\sigma_j^2+(\mu_i-\mu_j)^2)
\end{centering}
\end{equation}

\begin{equation}\label{eq:2c}
\begin{centering}
\sigma_{Final}^2=c \cdot \sigma_1^2 + (1-c) \cdot \sigma_2^2 +c \cdot (1-c) \cdot (\sigma_1^2 + \sigma_2^2 +(\mu_1-\mu_2)^2)
\end{centering}
\end{equation}

\subsubsection*{PSF as a function of lens aperture} 

\begin{figure}[htb!]
    \centering
    \includegraphics[width=\textwidth]{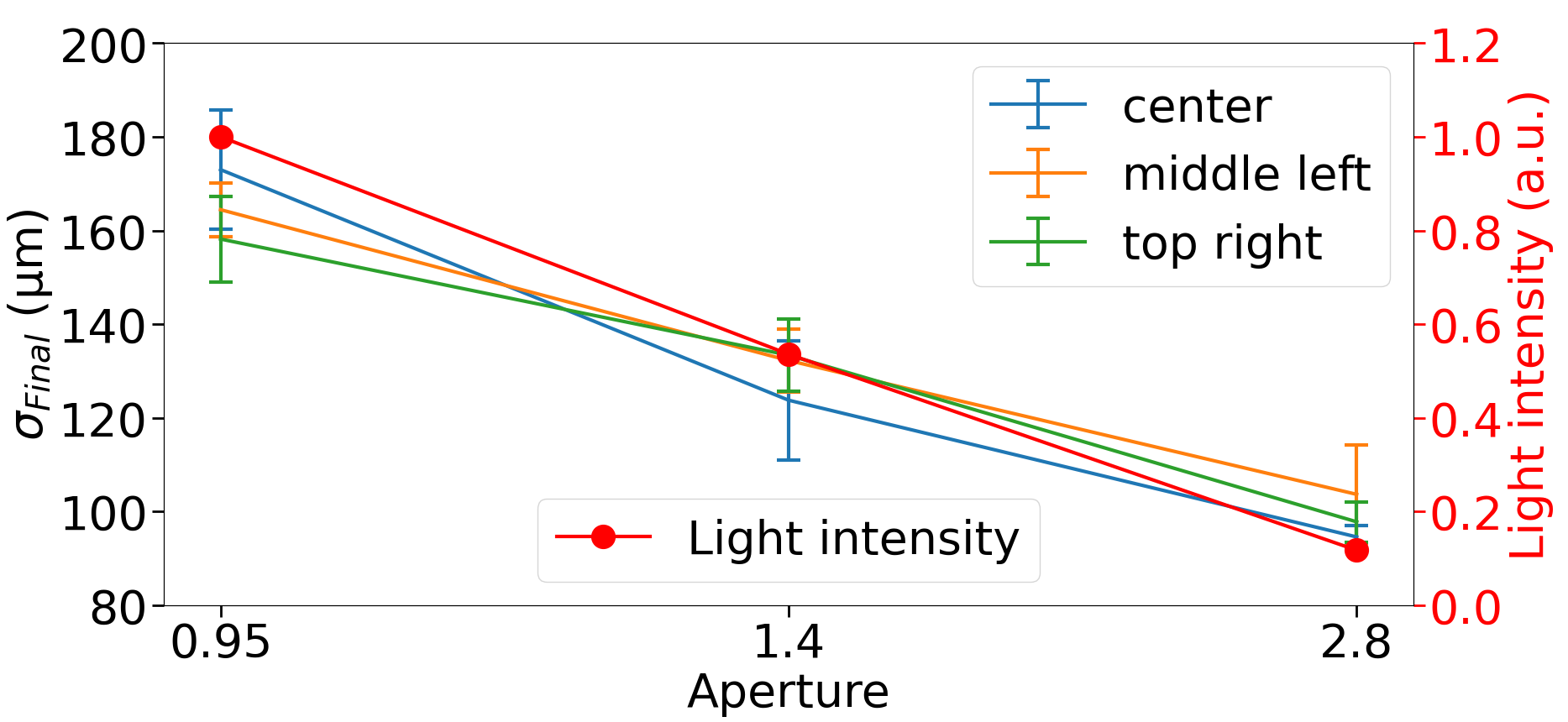}  
    \caption{Standard deviation extracted from the fit of the PSF shown in Figure\,\ref{Fig:PSFPositions} where $\sigma_{Final}$ is the average of the vertical and horizontal standard deviations described in \eqref{eq:2c}. Total light intensity (red line) that varies in $\frac{1}{(f/\#)^2}$ with  aperture $f/\#$. The lens aperture values f/0.95, f/1.4, and f/2.8 are tested for three beam positions. The drift field is 350\,V/cm and the amplification field is 47\,kV/cm.}
\label{Fig:ApertureScan1Dlines}
\end{figure}

The PSF was measured for three distinct lens aperture settings to evaluate their effect on the spatial resolution. Figure~\ref{Fig:ApertureScan1Dlines} illustrates the average standard deviations ($\sigma_{Final}$) of the horizontal and vertical profiles for three beam positions on the detector. The $\sigma_{Final}$ value decreases for larger f-numbers and remains almost unchanged for different beam locations. This shows the influence of the lens aperture on the signal spreading, decreasing by about 37\,\% for apertures from f/0.95 to f/2.8. While the shape of the aberrations depends on the beam position, it has almost no impact on the overall signal spreading.

\begin{figure}[htb!]
    \centering
    \includegraphics[width=0.6\textwidth]{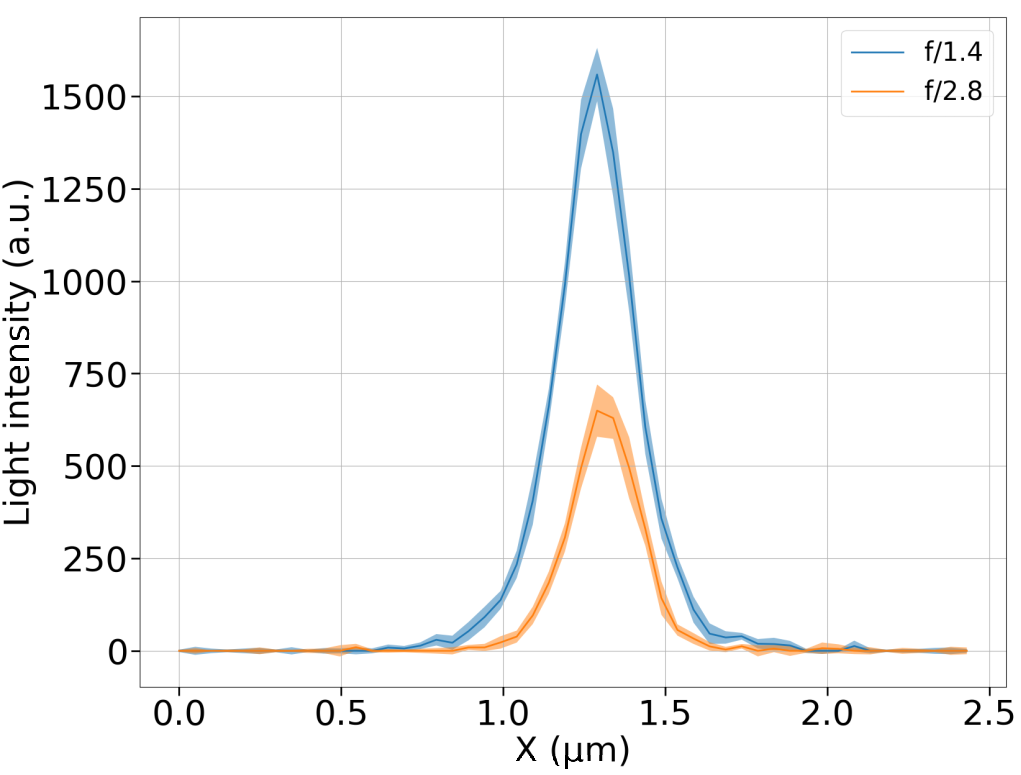}   
    \caption{1D intensity line profiles for a beta mesh, at a drift and amplification fields of 350\,V/cm and 47\,kV/cm, and lens apertures of f/1.4 and f/2.8.}
\label{Fig:1DaperturesCompare}
\end{figure}

%%%%Changer les p
As shown, lens aberrations significantly impact measurements at a lens aperture of f/1.4. To account for this, measurements under the same conditions (beta mesh, 350\,V/cm drift field and 47\,kV/cm amplification field) were taken with both f/1.4 and f/2.8 lens apertures. Assuming that aberrations are negligible at a lens aperture of f/2.8, we can express the aberration contribution ($\sigma_A^2$) as:
\begin{equation}
\sigma_A^2 = \sigma_{f/1.4}^2 - \sigma_{f/2.8}^2
\end{equation}
where $\sigma_{f/1.4}$ and $\sigma_{f/2.8}$ are the total measured standard deviations at lens apertures of f/1.4 and f/2.8, respectively. Figure~\ref{Fig:1DaperturesCompare} shows the intensity line profiles for both lens apertures, from which $\sigma_A = 81.0\pm2.3\,\microns$ was extracted.

\subsubsection*{PSF as a function of drift field} 
 %%%
 Different drift field values for 2\,mm and 4\,mm drift gaps were compared using a lens aperture of f/1.4 with a beta mesh.

\begin{figure}[htb!]
    \centering
    \includegraphics[width=\textwidth]{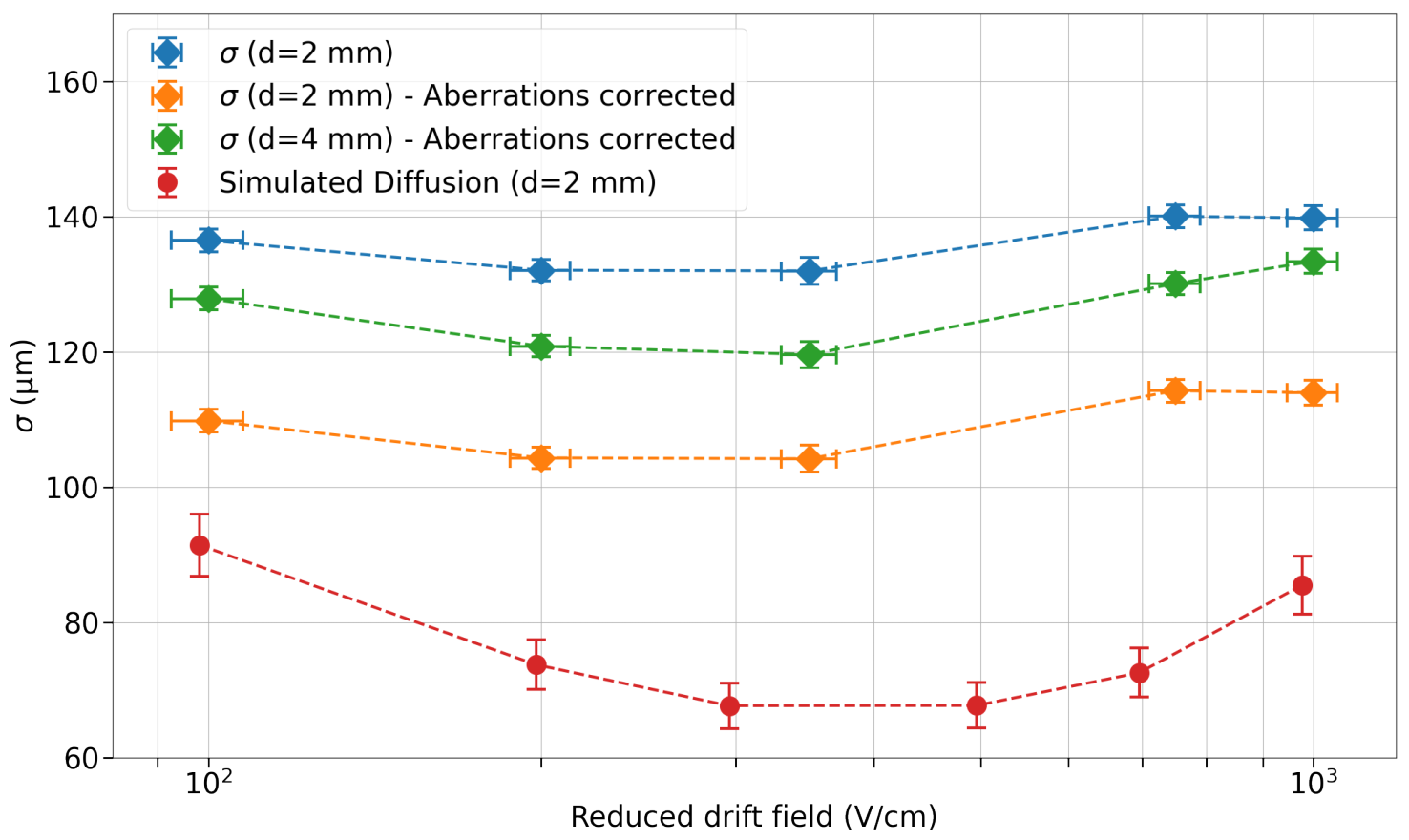}
    \caption{Dependence of the PSF width with the reduced drift field for a 2\,mm drift gap  (blue and orange markers) and 4\,mm drift gap  (green markers). The errors are computed by error propagation from the fit parameters uncertainty. Simulated diffusion standard deviation extracted from~\cite{BIAGI1999234} for a 2\,mm drift gap (red markers). The errors are computed from the uncertainty on the cathode planarity that brings an error on the drift length $d$, involved in the diffusion standard deviation: $\sigma_D \propto \sqrt{d}$. The optical aberrations have been corrected.}
\label{Fig:SigDriftScanCorrectedSimu_d2_4mm}
\end{figure}

 In Figure~\ref{Fig:SigDriftScanCorrectedSimu_d2_4mm}, the standard deviation ($\sigma$) is extracted from the PSF fit for five different drift field values, from 100\,V/cm to 1000\,V/cm, at an amplification field of 46\,kV/cm and for a lens aperture of f/1.4. The aberration contribution has been corrected as follows
 $\sigma^2=\sigma_{f/1.4}^2 - \sigma_A^2 $. 
Simulated diffusion on Magboltz~\cite{BIAGI1999234} values have been included in the plot. The measured values exhibit a similar trend, showing a minimum around 300\,V/cm. The difference between simulated and measured values can be attributed to the electron range  and to light reflections.

We have assumed that the PSF width is the quadratical sum of three different contributions:

\begin{equation}
\sigma^2= \sigma_{A}^2 + \sigma_{D}^2 +\sigma_{Res}^2 
\label{eq:res}
\end{equation}
where $\sigma_{A}^2$ is the contribution from aberrations, $\sigma_{D}^2$ quantifies the effect from electron diffusion, and $\sigma_{Res}^2$ encompasses the residual contribution arising from the electron range and lightreflections. From Equation~\ref{eq:res}, we can estimate $\sigma_{Res} = 75.5 \pm 4.3,\microns$.

%%%%%% A EXCLURE POUR l'instant en attendant resultat Apylux%%%%
%then the weights of the different contributions can be extracted and are given in table~\ref{tab:Weight-contributions}. Other minor (?) effects, like mesh reflections have not been taken into account.

 %\begin{table}[]
 %    \centering
  %   \begin{tabular}{|c|c|c|c|}
   %  \hline
    %      Drift gap& $\sigma_{A}$ ()& $\sigma_{D}$ ()& $\sigma_{R}$ ()\\
     %     \hline
      %   2\,mm & 36 & 32& 32\\
       %  \hline
        %  4\,mm& 32& 38& 30\\
         % \hline
     %\end{tabular}
     %\caption{Weight of the different estimated contributions to the total PSF coming from lens aberration ($\sigma_{A}$), gas diffusion ($\sigma_{D}$) and electron range ($\sigma_{R}$) for the 2\,mm and 4\,mm drift gaps with the Beta mesh and lens aperture f/1.4.}
     %\label{tab:Weight-contributions}
 %\end{table}
 
%%\end{figure}

%%
\subsubsection*{PSF as a function of mesh type} 
\begin{figure}[htb!]
    \centering
    \includegraphics[width=0.44\textwidth]{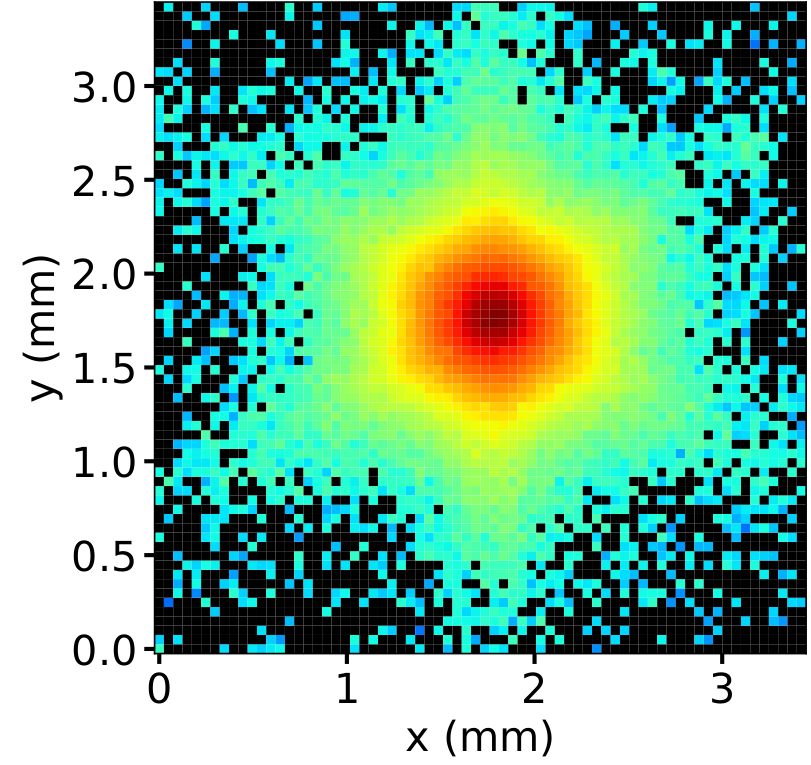}  
    \includegraphics[width=0.44\textwidth]{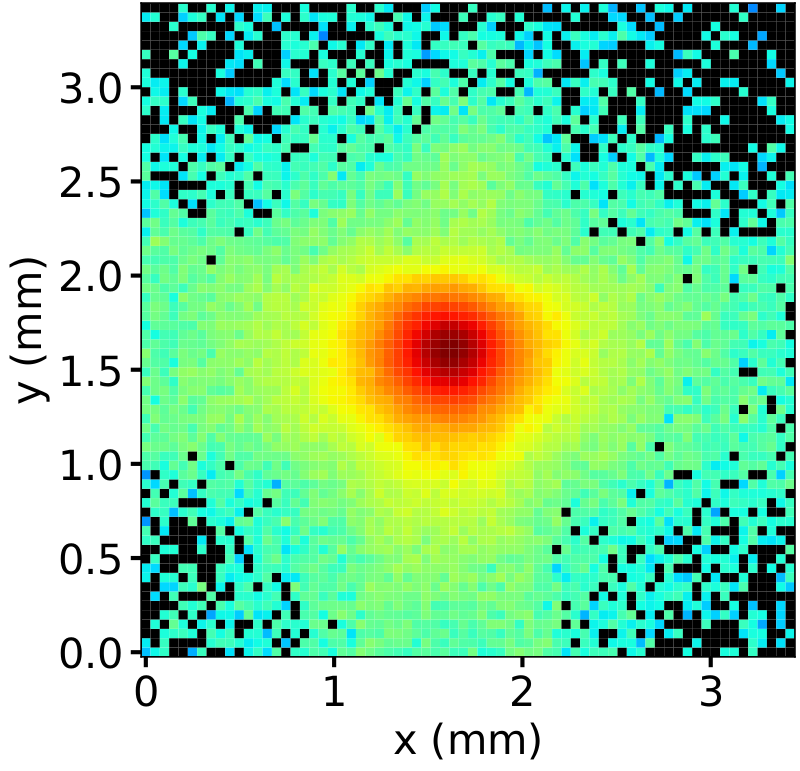}  
    \includegraphics[width=0.09\textwidth]{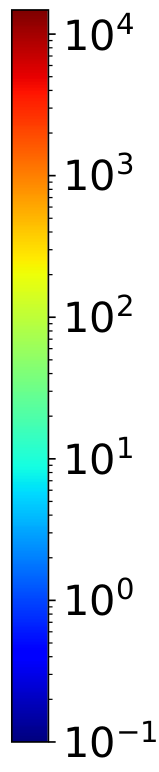}  
    \caption{2D profile of a point-like beam source in the center of the image plane for beta mesh (left) and standard mesh (right). The lens aperture is f/1.4, the drift field is 350\,V/cm. The amplification field is 59\,kV/cm with the beta mesh where the amplification gap is $75\,\microns$ (left), and 41\,kV/cm with the standard mesh where the gap is $128\,\microns$ (right). The color scale is logarithmic. Different reflection patterns are observed on the figures.}
\label{Fig:MeshReflection}
\end{figure}

Figure\,\ref{Fig:MeshReflection} shows PSF profiles with a beta and a standard mesh in high amplification field conditions (gain $\approx 3\times10^3$). Different reflection patterns can be observed that are correlated with the mesh geometry i.e. hexagonal-like shape for the beta mesh and a cross-like shape for the standard, woven mesh. 
Reflection patterns are produced by the shiny stainless steel during the avalanche multiplication, but they are not observed in images at low X-ray flux or at a low detector gain. We estimate that their intensity is two orders of magnitude lower than the central light signal.

%The mesh reflects back the light that is emitted isotropically during avalanche multiplication. Even if the mesh is made of shiny stainless steel, this reflection has been observed only with large X-ray flux and large detector gain, when light intensity is sufficient to see this effect that is around 2 orders of magnitude lower than the central light signal.
Reflection patterns are produced by the shiny stainless steel during the avalanche multiplication, but they are not observed in images at low X-ray flux or low detector gain. We estimate that their intensity is two orders of magnitude lower than the central light signal at low detector gain.

\subsubsection*{PSF as a function of energy} 
\begin{figure}[htb!]
    \centering
    \includegraphics[width=0.22\textwidth]{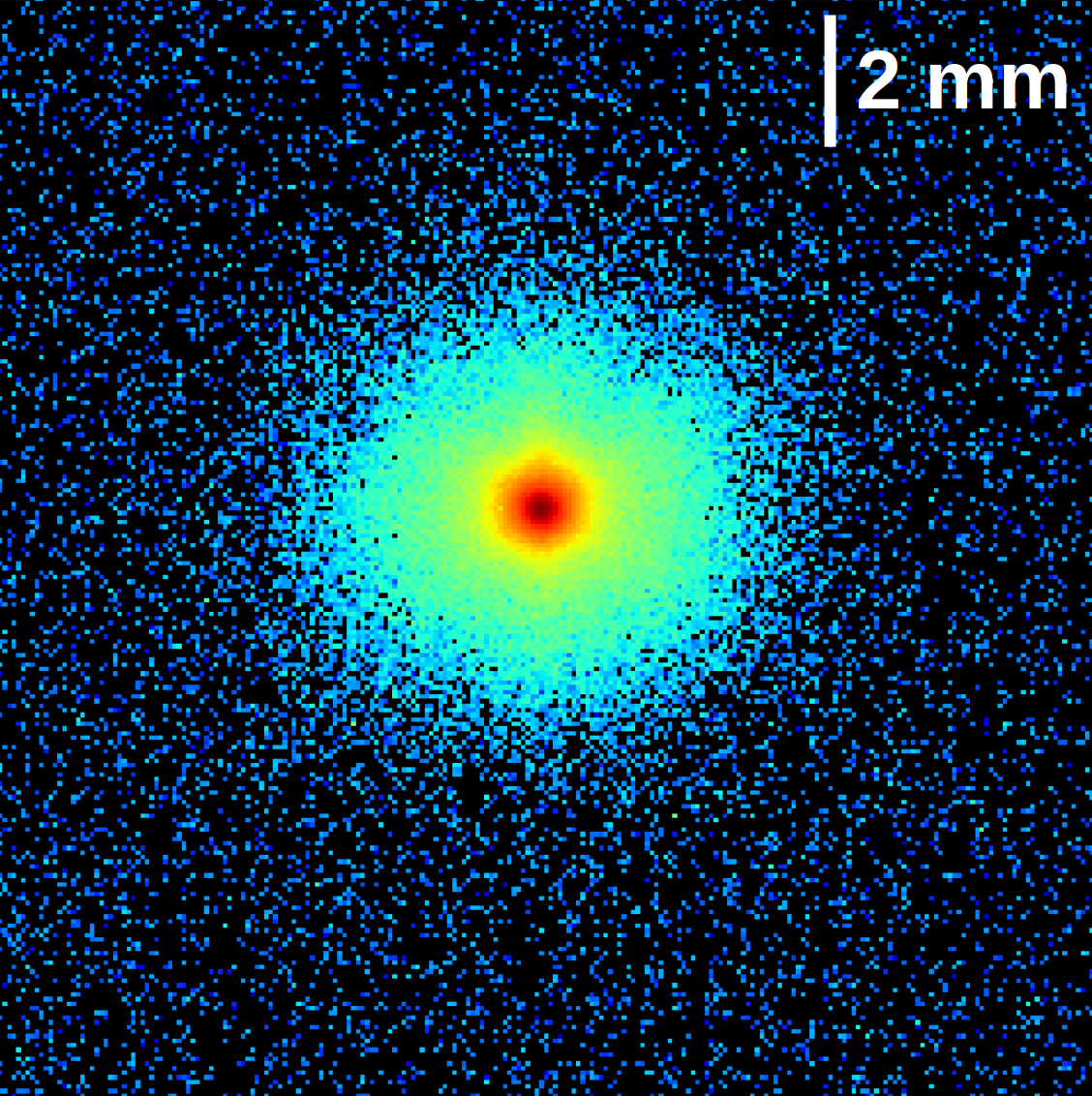}
    \includegraphics[width=0.22\textwidth]{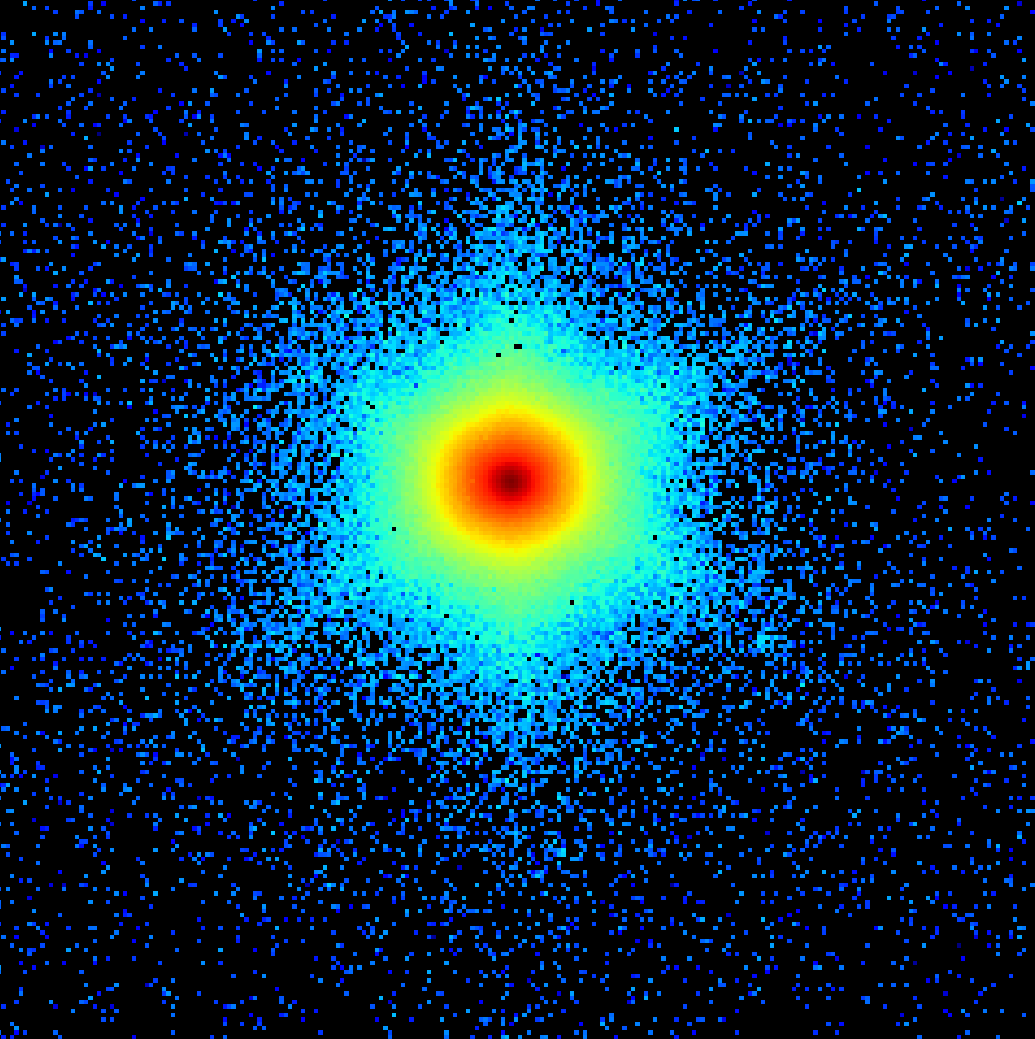}
    \includegraphics[width=0.22\textwidth]{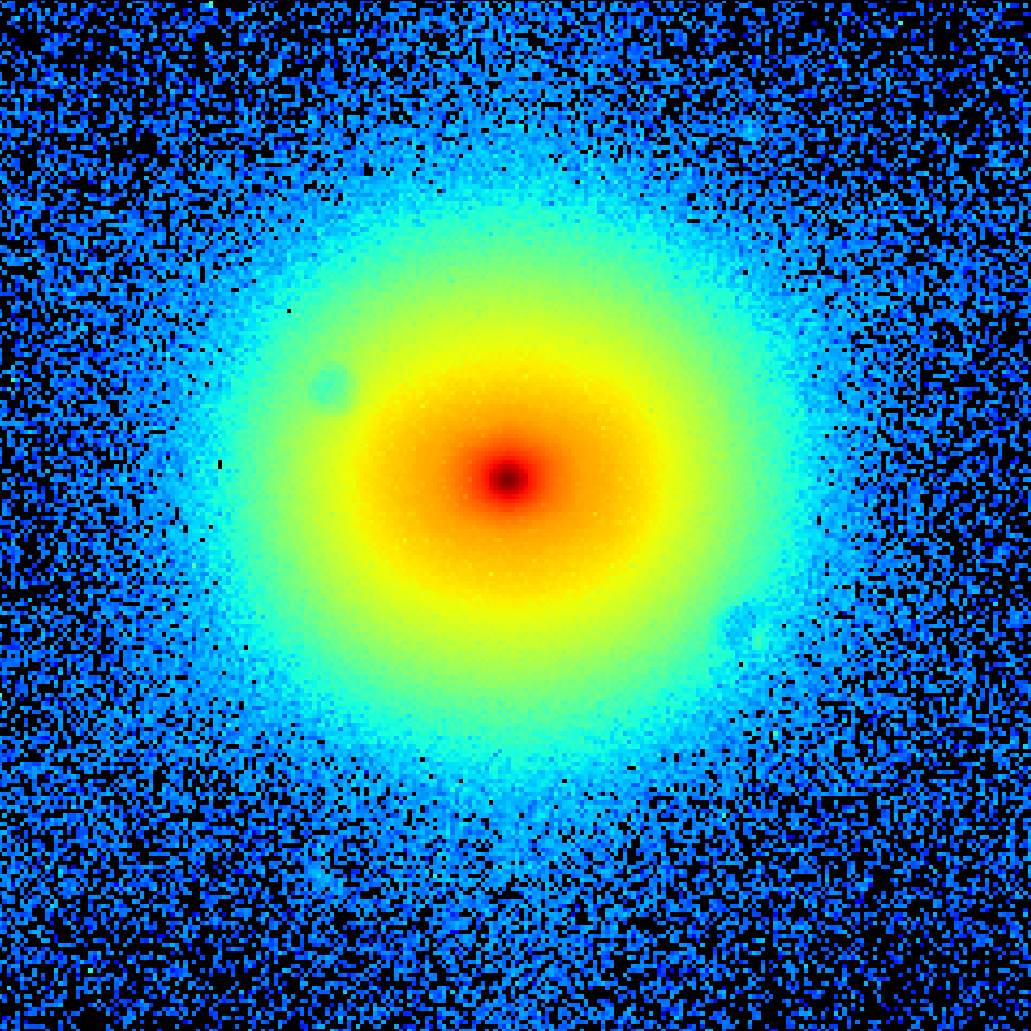}
   \includegraphics[width=0.22\textwidth]{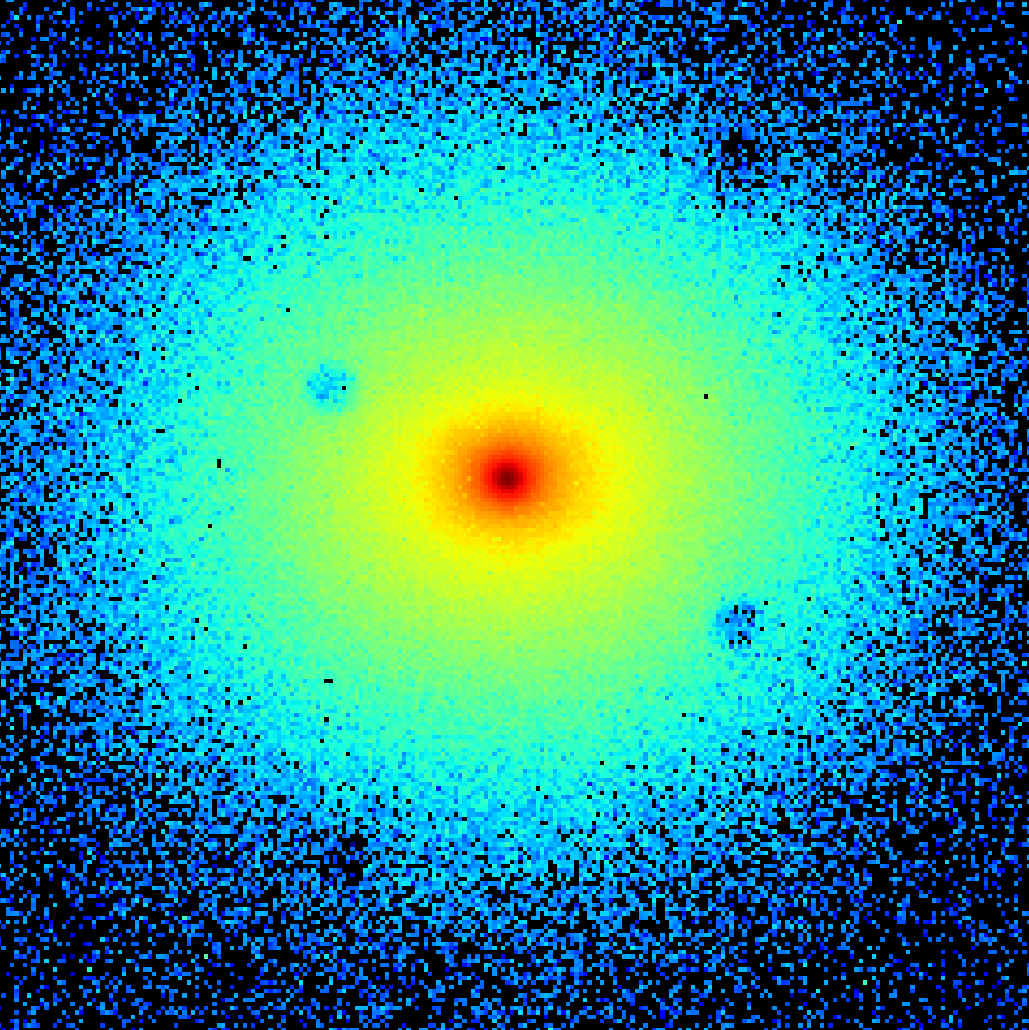}
    \includegraphics[width=\textwidth]{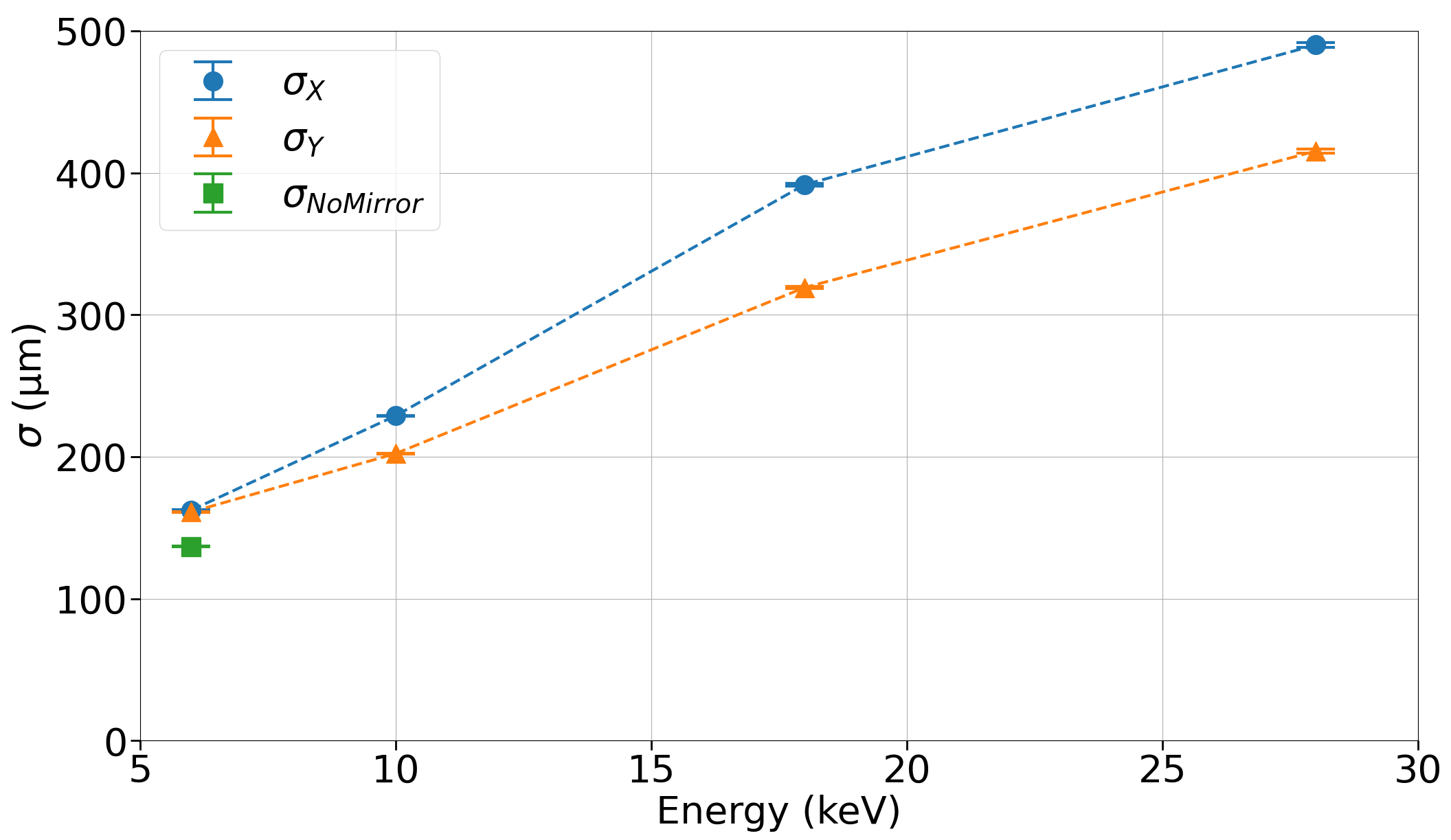}
     \caption{2D profile of the X-ray source at beam energies of 6\,keV,10\,keV, 18\,keV and 28\,keV from left to right. Standard deviations extracted from signal fit for four different X-ray beam energies in the horizontal ($\sigma_X$) and vertical ($\sigma_Y$) orientations (bottom) after correction of the diffusion contribution. The lens has a 25\,mm focal length, a mirror with a 90$^{\circ}$ orientation is used and the amplification and drift fields are 59\,V/cm and 350\,V/cm, respectively. Standard deviation with a 50\,mm focal length lens without mirror ($\sigma_{\mathrm{NoMirror}}$).}
\label{Fig:PSFEnergyScanBeta4mmMirror}
\end{figure}

Measurements were recorded at various X-ray energies up to 28\,keV with a 4\,mm drift gap (instead of 2\,mm) to maximize photon conversion and limit border effects. A 25\,mm focal length lens with an f/1.4 aperture was used. In order to protect the camera from high X-ray energies a mirror was included in the setup.  Figure~\ref{Fig:PSFEnergyScanBeta4mmMirror} shows the PSF measured with and without the mirror at 6\,keV. The PSF values are comparable, a small degradation of less that 30\,\microns is observed.

Figure \ref{Fig:PSFEnergyScanBeta4mmMirror}-top shows the 2D profile profiles at X-ray beam energies of 6, 8, 18, and 28\,keV showing a wider spot with increasing energy due to the larger electron range.

The contribution from the electron diffusion, simulated with Magboltz\cite{BIAGI1999234} at a drift field of 350\,V/cm,  was quadratically substracted from the PSF width and is shown in Figure~\ref{Fig:PSFEnergyScanBeta4mmMirror} (bottom). The signal width increases with beam energy by approximately 14\,\microns/keV, with the horizontal width being on average 15\% larger than the vertical width.

\subsection{PSF measurements with GEM}

PSF width recorded with different GEM geometries was measured in a comparable configuration as used for Micromegas.

\begin{figure}[htb!]
    \centering
    \includegraphics[width=0.45\textwidth]{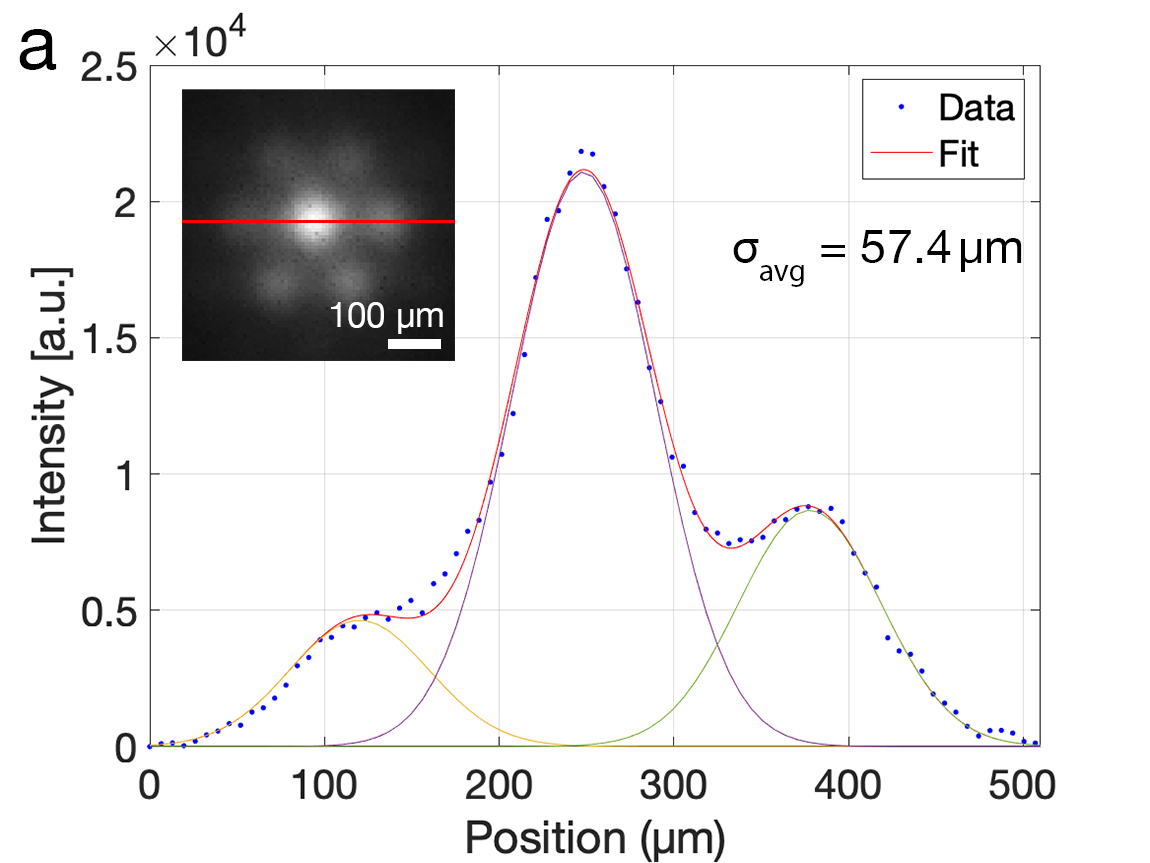}
\includegraphics[width=0.45\textwidth]{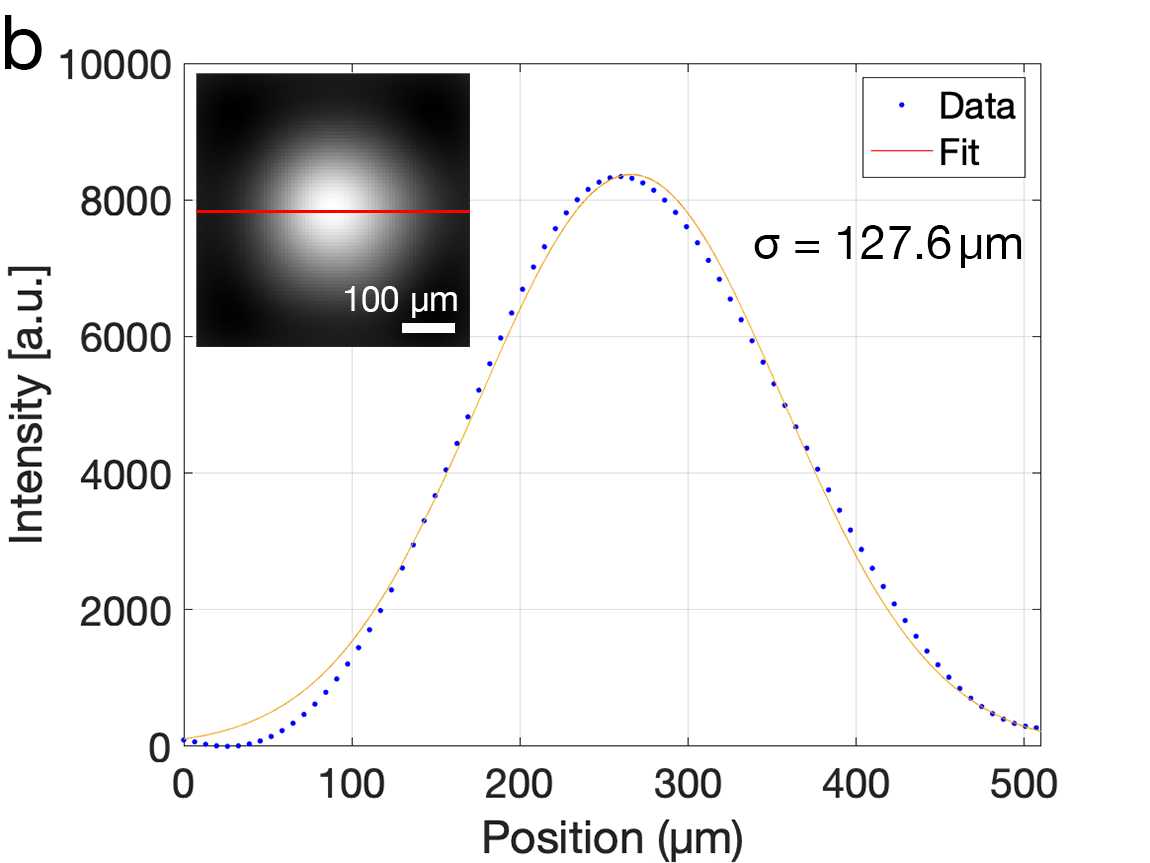}
   \caption{Signal from standard thin GEM: a: Line profile and fit with three Gaussians for adjacent holes. b: Line profile after FFT and its corresponding Gaussian fit. Insets: images indicating line profiles across GEM holes.}
    \label{GEMFitFFT}
\end{figure}

Signals of a 6\,keV X-ray beam with a size of 20\,$\times$ 20\,$\microns^2$ on a hole of the GEMs were recorded to evaluate the PSF of the detectors. The response on a single standard thin GEM foil is shown in Figure~\ref{GEMFitFFT}a. The central hole and one ring of neighboring holes are contributing to the observed signal intensity. A horizontal line profile of pixel value intensity was taken and three Gaussian functions were used to fit the signal contributions of the individual GEM holes. The position of the Gaussian was fixed according to the nominal position of GEM hole centers.

A mean width of $\sigma \approx 57 \mu$m was determined for the signal contribution from each hole for single thin GEM foils by fitting line profiles with multiple Gaussian functions. The fit uncertainty of each Gaussian function was $\approx 10 \%$. 

To quantify the PSF width, the contribution to the signal from three GEM holes (central one and two neighboring holes) has been considered and a Fast Fourier Transform (FFT) of the original image was used to remove signal frequency components corresponding to the discrete hole pattern. The inverse FFT image after removal of signal frequencies corresponding to the hole pattern is shown in Figure~\ref{GEMFitFFT}b. A Gaussian fit of the resulting image yields a PSF width of 
$\sigma = 127 \microns$ which represents the PSF of the detector taking into account the sharing of the signal into neighboring GEM holes.

%\begin{figure}
 %   \centering
    %\includegraphics[width=0.9\textwidth]{figures/StdGEM_diffusionVsDriftField.png}
   %\caption{Minimum diffusion component in X and Y directions recorded with thin standard GEM determined with 2D Gaussian fit. (\EFR{To be removed from here and merged into comparative plot with MM}}
    %\label{GEMDiffusionComponent}

%\end{figure}

\begin{figure}[htb!]
    \centering
    \includegraphics[width=0.5 \textwidth]{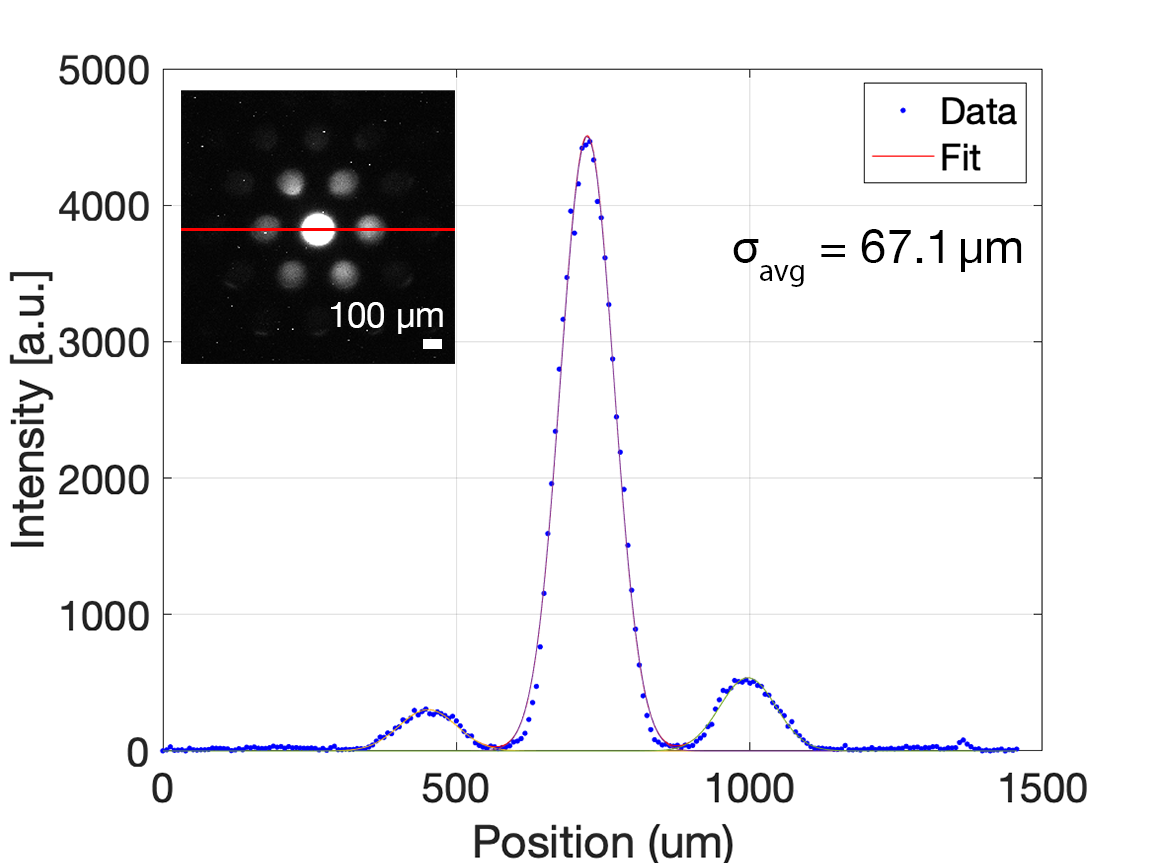}
   \caption{Signal from glass GEM: Line profile and fit with three Gaussian functions for adjacent holes. Inset: image indicating line profile across GEM hole.}
    \label{GlassGEMFitFFT}
\end{figure}

%For the glass GEM with a larger hole pitch and diameter, comparable images were acquired in the same configuration as used for the standard thin GEM. A comparable width of $\sigma \approx 67 \microns$ of the response from a single hole was extracted from a Gaussian fit to a line profile as shown in Figure \ref{GlassGEMFitFFT}. Despite the larger hole diameter of $160\,\microns$ compared to $70\,\microns$ diameter holes of the standard thin GEMs, the observed light profile of glass GEMs does not appear significantly wider.
For the glass GEM, beam images appear similar to those of standard thin GEMs. Indeed, a similar PSF width value of $\sigma \approx 67 \microns$ is extracted, as shown in Figure~\ref{GlassGEMFitFFT}, even through glass GEM holes are significantly wider (diameter of $160\,\microns$ compared to $70\,\microns$). This suggests that a significant part of the avalanche multiplication occurs in the center of the hole. Scintillation light emitted during avalanche multiplication in the holes may be collimated by the long glass GEM holes. In fact, compared to the $50\,\microns$ substrate thickness of standard thin GEMs, glass GEMs are more than a factor 10 thicker with a substrate thickness of $570\,\microns$. Furthermore, the cylindrical glass GEM holes may have a stronger collimating effect than the double-conical thin GEM holes. 
The increased pitch of the glass GEM holes combined with a similar width of light emission results in a clear separation of holes as shown in Figure~\ref{GlassGEMFitFFT}. To estimate the spread of signals due to collection into the glass GEM, a Gaussian envelope containing the response of neighboring holes for a beam aligned with the central hole was used and a width of $\sigma = 200\,\microns$ was extracted. Compared to the Gaussian envelope width of $\sigma = 127 \,\microns$ determined for a thin GEM, the increased width reflects the larger hole pitch of the glass GEM.

In multiplying structures with features above a certain dimension, electron avalanches may preserve information of the location of incident primary charges and asymmetric signal intensity is observed for primary charges which do not enter the holes axially centered. This behavior was observed for THGEMs and THCOBRA detectors and confirmed by simulations~\cite{Garcia_2021}.
In structures with small holes such as standard thin GEMs, this effect is not observed and avalanche multiplication appears to occur symmetrically across the hole cross-section.

\begin{figure}[htb!]
    \centering
    \includegraphics[width=0.7\textwidth]{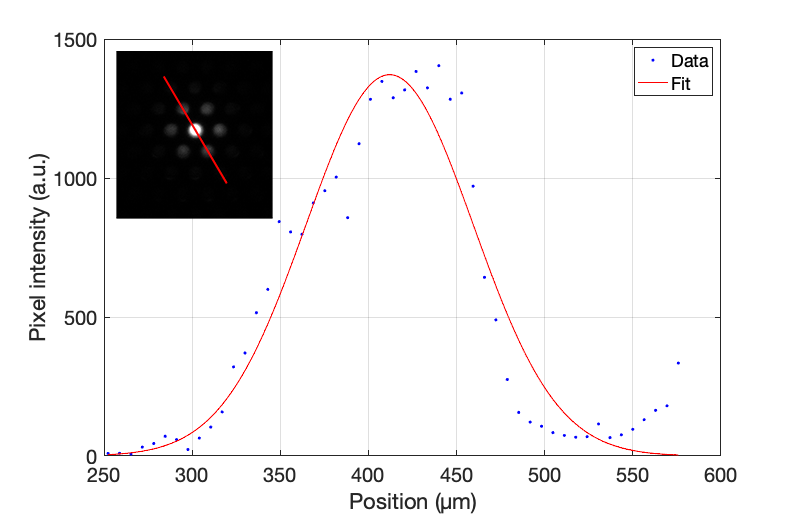}
   \caption{The asymmetric line profile across a glass GEM hole points to retaining of position information during avalanche multiplication. Inset: image indicating line profile of neighbouring hole with beam centered on central hole.}
    \label{GlassGEMAsymmetry}
\end{figure}

The preservation of location information may be used to improve the spatial resolution achievable with structures featuring relatively large hole sizes and pitches. The response of standard thin GEM and glass GEM detectors was examined for possible asymmetric signal intensity within holes. A line profile across a hole adjacent to the central hole with highest intensity is shown in Figure~\ref{GlassGEMAsymmetry}. In this case, primary charges are centered at the brightest hole and the line profile of the neighboring hole exhibits an asymmetric shape with enhanced signal intensity towards the direction of the hole on which the beam was centered. 

%Asymmetric pixel value profile across hole in glass GEM shows retaining of position information during avalanche multiplication. 

\subsection{Beam position measurement with GEM}

\begin{figure}[htb!]
    \centering
    \includegraphics[width=0.9\textwidth]{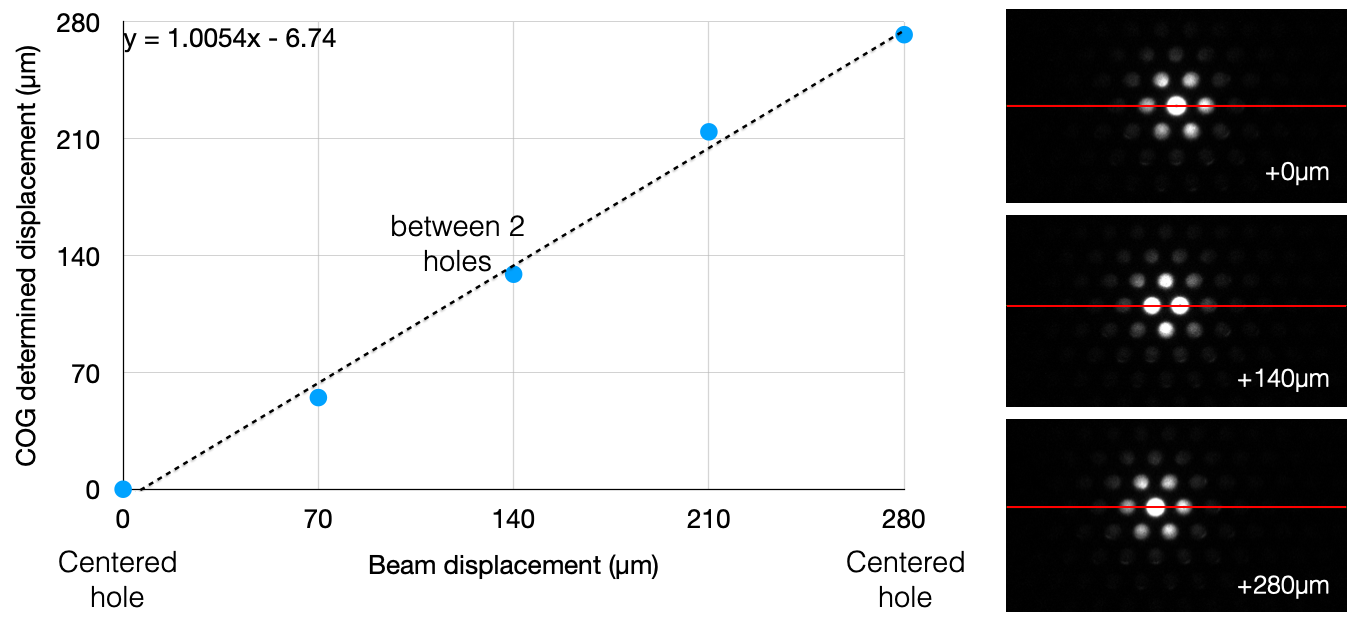}
   \caption{Left: Correlation of reconstructed and nominal beam spot position for displacements from 0\,$\microns$ to 280\,$\microns$ recorded with glass GEM. Right: Exemplary images at three different displacements showing contribution of neighbouring GEM holes.}
    \label{glassGEMBeamPositionReconstruction}
\end{figure}

 The achievable resolution in determining the beam position was compared for the two GEM geometries. The position of an incident beam was determined by the center-of-gravity of the image after background subtraction. Incident beam position was changed by displacing the detector using the movable stage on which it was mounted. Reconstructed and nominal position were compared for displacements up to one hole pitch of the detector. A correlation of reconstructed and nominal beam position and exemplary images of three beam positions recorded with a glass GEM are shown in Figure~\ref{glassGEMBeamPositionReconstruction}.

A linear fit of the reconstructed beam position as a function of nominal beam position was performed and the mean residual of the data points with respect to the fit was used to quantify the localisation resolution. A mean residual of $\sigma_{mean} = 9.5 \pm 4.7 \,\microns$ was achieved with the glass GEM. For the standard thin GEM, a mean residual of $\sigma_{mean} = 1.6 \pm 0.5 \,\microns$ was obtained. The preservation of spatial information during avalanche multiplication in the glass GEM may contribute to the good reconstruction accuracy obtained with this amplification structure. 

Signals shared between multiple GEM holes can be used to determine the position of incident beams with a resolution superior to the pitch of the GEM. The localisation resolution achieved with glass and thin GEM using a center-of-gravity method are significantly below the hole pitch (280 and 140\,$\microns$, respectively) of the structures and demonstrates the possibility of accurate beam position determination with these amplification structures.

\subsection{Micromegas pillars effects}
\begin{figure}[htb!]
    \centering
    \includegraphics[width=0.3\textwidth]{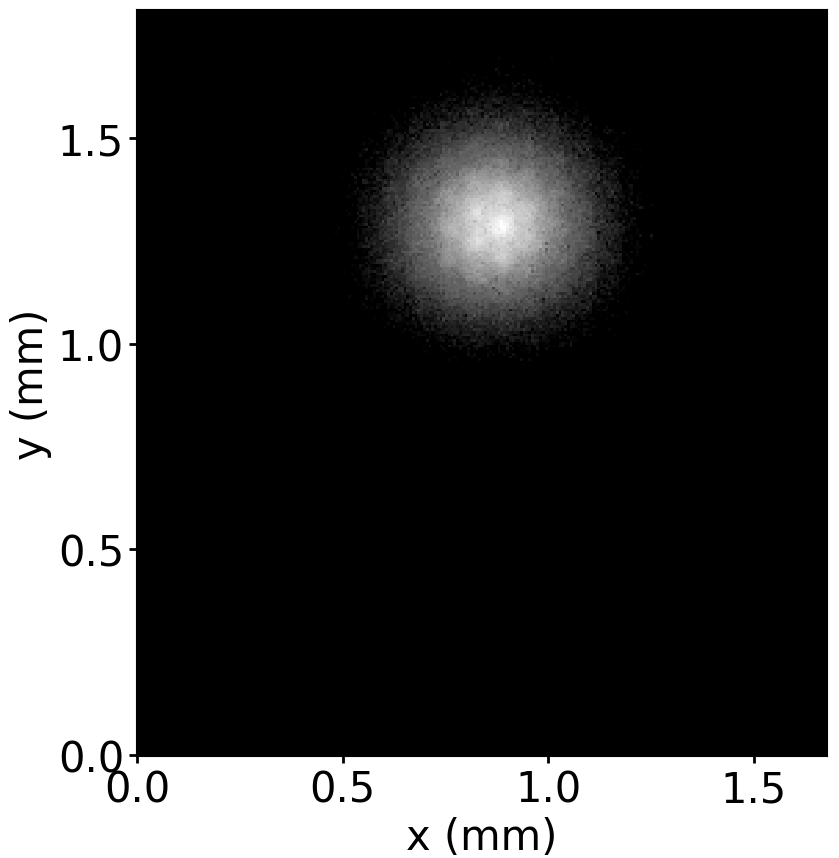}
    \includegraphics[width=0.3\textwidth]{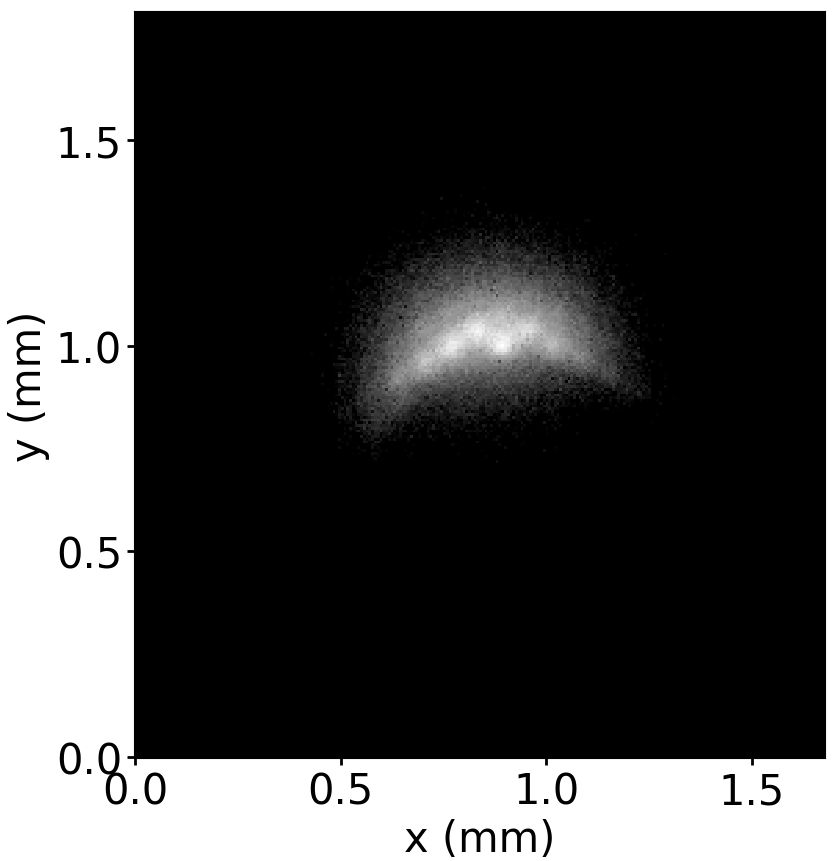}
    \includegraphics[width=0.3\textwidth]{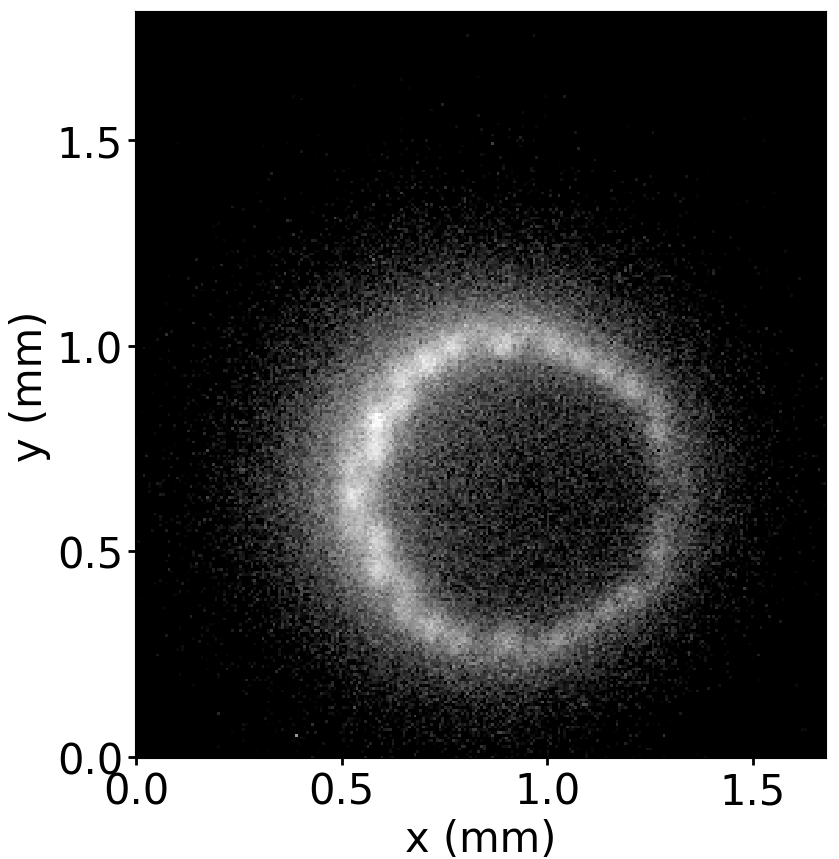}
    \includegraphics[width=0.048\textwidth]{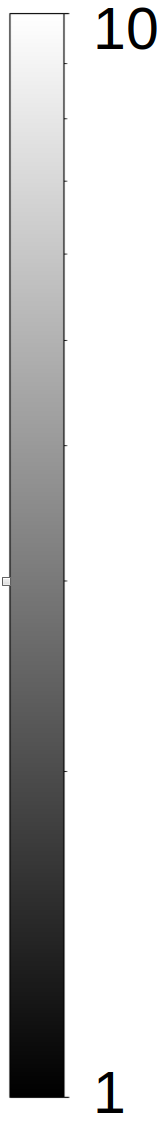}
    \includegraphics[width=\textwidth]{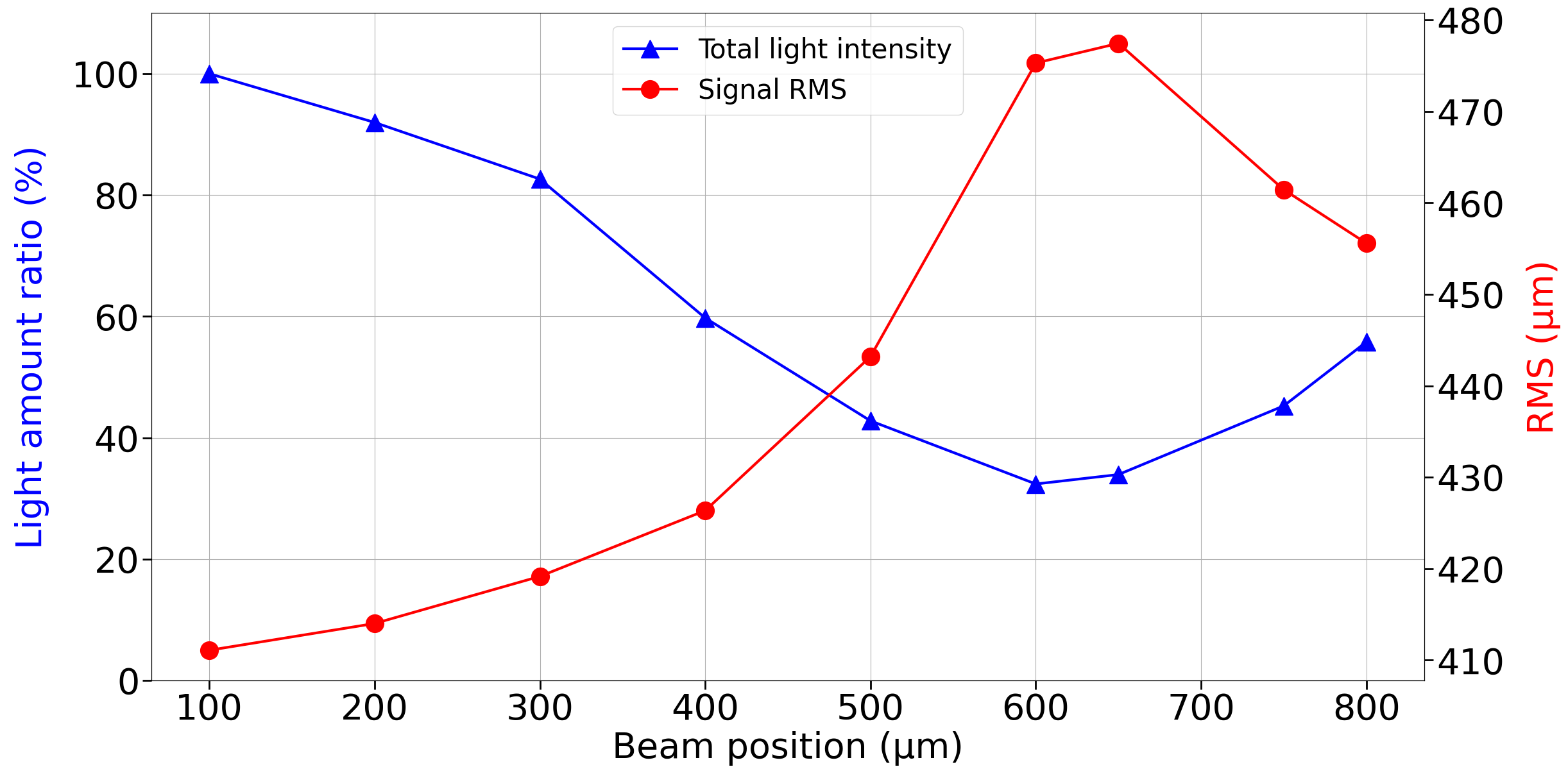}
    \caption{Beam profile at different positions $Y$ of the detector. Tthe gray scale is logarithmic and the same scale is used for all images. The \textit{$X$} position is at 800\,\microns for the three plots. Top left: beam is out of the pillar ($Y=100\,\microns$). Top center: the pillar covers a beam section ($Y=400\,\microns$). Top right: beam is fully covered by the pillar ($Y=600\,\microns$)  Bottom: normalized total signal amount (blue curve) and the total signal RMS (red curve). 30\,\% of the light amount remains when the beam and a pillar overlap and light signal is enlarged by 15\,\%.}
\label{Fig:AcrossPillar}
\end{figure}

A thin beam ($20 \times 20\,\microns^2$ size, 6\,keV X-ray energy) with
with a 50\,mm focal length lens to study light emission near the detector pillars. The detector was positioned at various heights, and the resulting beam profiles are shown in Figure~\ref{Fig:AcrossPillar}-top. Even with full overlap of the beam and the pillar (top-right image), a significant fraction of the signal remains, creating an eclipse-like shape. This effect is quantified by the total light intensity at different beam position (blue line in Figure~\ref{Fig:AcrossPillar}). The spreading caused by electron range and diffusion in a 2\,mm drift gap extends the signal by approximately 100\,\microns. Therefore, beyond a diameter of $4\sigma$ (400\,\microns), we would expect that less than 5\% of the signal remains for a Gaussian beam profile. However, the total light intensity is 30\% of the full beam light when the beam is completely covered by the pillar (Z=600\,\microns).

 This fact is explained by the maximum of the electrons drifting towards a pillar are being deflected towards the amplification gap due to the curvature of the field lines around the pillar. This fact is supported by the maximum of the beam signal RMS (red line) when the beam overlaps with the pillar, indicating that the enlargement is caused by the pillar rather than typical effects such as electron range, diffusion, light reflection, or aberration. The curvature of the field lines around a dielectric spacer is a known effect and has been observed previously~\cite{BHATTACHARYA201541,Kuger:2017oqa}.

%thesis: https://cds.cern.ch/record/2277011/files/CERN-THESIS-2017-106.pdf
%cited here paper cited in thesis, can include both references

Figure~\ref{Fig:AcrossPillar} shows a hexagonal structure, representative of the mesh structure shown in Figure~\ref{Fig:MPGDGeometries}. As expected, the  projected amplified electron cloud  from the ionisation of the gas in the drift gap is made up of multiple sources from the electrons crossing the mesh reaching the amplification zone.  So, the PSF represents the accumulated photons  from electron avalanches with a mesh pattern. The same is true for the GEM given in Figure~\ref{Fig:SOLEILmmVSgem}. 
The photo-electron track width has been estimated using a Geant4~\cite{Agostinelli:2002hh} simulation. The extracted width is $\simeq$35\microns.  Comparing this width with extracted PSF taking into account diffusion, shows that the projected amplified cloud is consistent with the simulation.

\subsection{Comparison}
\begin{figure}
    \centering
\includegraphics[width=0.9\textwidth]{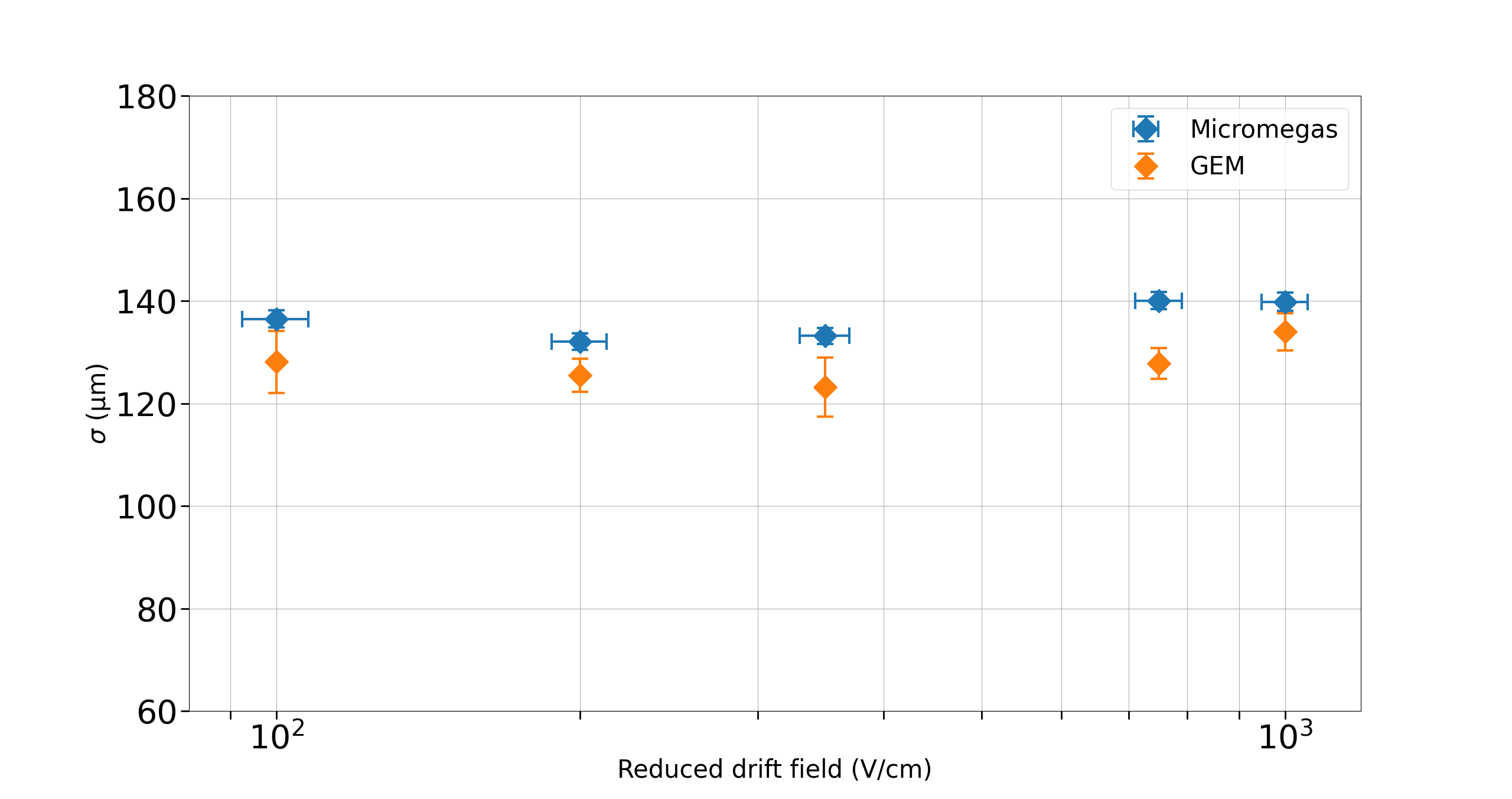}
   \caption{Distribution among five drift field values of the extracted standard deviation from the fit for a 2\,mm drift gap  with a standard mesh Micromegas detector (blue markers) and with a single standard GEM detector (orange markers). The lens has a 25\,mm focal distance and the aperture is f/1.4. The gain is about 500 for both detectors.}
    \label{Fig:SOLEILmmVSgem}
\end{figure}

Images recorded with Micromegas and with a single standard thin GEM were fitted by 2D Gaussian to extract the PSF width for varying drift fields for 2\,mm drift region in the same conditions. The lens has a 25\,mm focal distance and the aperture is f/1.4. The gain is about 500 for both detectors. The dependence of PSF width on the drift field  is shown in Figure \ref{Fig:SOLEILmmVSgem}.
The trend, for both Micromegas and GEM, is of the same order and shows a minimum at the expected minimum of diffusion ($\approx$ 350\,V/cm) in agreement with the simulated diffusion shown in Figure~\ref{Fig:SigDriftScanCorrectedSimu_d2_4mm}

%%%%%%%%%%%%%%%%%%%%%%%%%%%%%%%%%%%%%%%%%%%%%%%%%%%%%%
\section{Conclusions and perspectives}
\label{section:Conclusion}
%%%%%%%%%%%%%%%%%%%%%%%%%%%%%%%%%%%%%%%%%%%%%%%%%%%%%%
 The use of a small incident X-ray beam allowed an accurate visualisation of the detector response to a point-like source. Measurements of the PSF of optically read out Micromegas and GEMs are used to compare the achievable spatial resolution of these MPGDs.Differences in the light emission profile between Micromegas and GEMs were observed and are attributed to their specific amplification structures.

The width of the detector response to a point-like source is determined by a combination of electronic and optical effects. Following interaction in the drift region, primary electron range and transverse diffusion as well as collection into the amplification structure contribute to a widening of the charge distribution. In an optical readout configuration, optical effects including reflections and lens aberrations further widen the observed signal.

Smaller apertures were effective in minimising the effect of lens aberrations in conditions where light intensity is sufficiently high. Reflections from the amplification structure were observed which depend on the geometry of the camera-facing electrodes. Anti-reflective layers or absorbing coatings could be effective in minimising reflections and improve overall image definition and thus spatial resolution. In addition to thin layer anti-reflective coatings, oxide surfaces or diamond-like carbon (DLC) coatings may be used to render electrodes non-reflective. Efforts in this direction are in progress.

Due to their larger hole pitch, GEMs showed an overall larger PSF  ($\approx$127\,$\microns$) when taking into account the collection of signals into multiple neighboring holes compared to the finer pitch of Micromegas meshes enabling to reach PSF of $\approx$108\,$\microns$. Nevertheless the difference is relatively small.  Work to explore other measures, such as Modulation Transfer Function (MTF) will give a broader comparison of the different amplification structures.

For larger GEM hole diameters, an asymmetry in the response of off-center holes was observed which can be attributed to a preservation of spatial information of the incident charge during the amplification process. This effect was seen for glass GEMs but not for standard thin GEMs suggesting that avalanches in holes below a certain size, electron avalanches fill the full hole. Good beam position reconstruction accuracy was observed for GEMs with different hole pitches demonstrating that recorded light intensity can be used as weighting factor.

The studies of PSF and microscopic images of the light emission profiles compare the response of MPGD technologies used with the optical readout approach. PSF can be used to deconvolve images and enhance image sharpness. Optimisations of amplification structures as well as the optical readout systems may be pursued to further increase the achievable spatial resolution and minimise distortions for applications requiring high imaging resolution or accurate track reconstruction.

%%%%%%%%%%%%%%%%%%%%%%%
\section*{Acknowledgments}
%%%%%%%%%%%%%%%%%%%%%%%
The authors acknowledge the financial support of the Cross-Disciplinary Program on
Instrumentation and Detection of CEA, the French Alternative Energies and Atomic Energy
Commission, the P2I Department of Paris-Saclay University and the P2IO LabEx (ANR-10-LABX-0038) in the framework ‘Investissements d’Avenir’ (ANR-11-IDEX-0003-01) managed by the Agence Nationale de la Recherche (ANR, France). The authors acknowledge SOLEIL for provision of synchrotron radiation facilities (proposal number 20221337) and we would like to thank Pascal Mercere and Paulo Da Silva for assistance in using METROLOGIE beamline.

\bibliography{mybibfile}
\bibliographystyle{elsarticle-num}
\end{document}